\documentclass[twocolumn]{aastex631}
\usepackage{amsmath} % or simply amstext

\usepackage{comment}
\usepackage{outlines}
\usepackage{color}

\usepackage{appendix}
\usepackage{multirow}

\begin{document}

\title{Cross-correlation of Luminous Red Galaxies with ML-selected AGN in HSC-SSP: \\Unobscured AGN residing in more massive halos}
\shorttitle{Unobscured AGN residing in more massive halos}
\shortauthors{C\'ordova~Rosado et al.}

%% Note that the corresponding author command and emails has to come
%% before everything else. Also place all the emails in the \email
%% command instead of using multiple \email calls.
\author[0000-0002-7967-7676]{Rodrigo~C\'ordova~Rosado}
\affiliation{Department of Astrophysical Sciences, Peyton Hall, Princeton University, 4 Ivy Lane, Princeton, NJ 08544, USA}
\author[0000-0003-4700-663X]{Andy~D.~Goulding}
\affiliation{Department of Astrophysical Sciences, Peyton Hall, Princeton University, 4 Ivy Lane, Princeton, NJ 08544, USA}
\author[0000-0002-5612-3427]{Jenny~E.~Greene}
\affiliation{Department of Astrophysical Sciences, Peyton Hall, Princeton University, 4 Ivy Lane, Princeton, NJ 08544, USA}

\author[0000-0001-6941-8411]{Grayson~C.~Petter}
\affiliation{Department of Physics and Astronomy, Dartmouth College, 6127 Wilder Laboratory, Hanover, NH 03755, USA}

\author[0000-0003-1468-9526]{Ryan~C.~Hickox}
\affiliation{Department of Physics and Astronomy, Dartmouth College, 6127 Wilder Laboratory, Hanover, NH 03755, USA}

\author[0000-0002-5808-4708]{Nickolas~Kokron}
\affiliation{Department of Astrophysical Sciences, Peyton Hall, Princeton University, 4 Ivy Lane, Princeton, NJ 08544, USA}

\author[0000-0002-0106-7755]{Michael~A.~Strauss}
\affiliation{Department of Astrophysical Sciences, Peyton Hall, Princeton University, 4 Ivy Lane, Princeton, NJ 08544, USA}

\author[0000-0002-5870-6108]{Jahmour~J.~Givans}
\affiliation{Center for Computational Astrophysics, Flatiron Institute, 162 5th Ave, New York, NY 10010, USA}
\affiliation{Department of Astrophysical Sciences, Peyton Hall, Princeton University, 4 Ivy Lane, Princeton, NJ 08544, USA}

\author[0000-0002-3531-7863]{Yoshiki~Toba}
\affiliation{National Astronomical Observatory of Japan, 2-21-1 Osawa, Mitaka, Tokyo 181-8588, Japan}
\affiliation{Academia Sinica Institute of Astronomy and Astrophysics, 11F of Astronomy-Mathematics Building, AS/NTU, No.1, Section 4, Roosevelt Road, Taipei 10617, Taiwan}
\affiliation{Research Center for Space and Cosmic Evolution, Ehime University, 2-5 Bunkyo-cho, Matsuyama, Ehime 790-8577, Japan}

\author[0000-0003-4684-608X]{Cassandra~Starr~Henderson}
\affiliation{Scripps Institution of Oceanography, La Jolla, CA, USA}

\correspondingauthor{Rodrigo~C\'ordova~Rosado}
\email{rodrigoc@princeton.edu}

\begin{abstract}
Active galactic nuclei (AGN) are the signposts of black hole growth, and likely play an important role in galaxy evolution. An outstanding question is whether AGN of different spectral types indicate different evolutionary stages in the coevolution of black holes and galaxies. We present the angular correlation function between an AGN sample selected from the Hyper Suprime Camera Subaru Strategic Program (HSC-SSP) optical + Wide-field Infrared Survey Explorer (WISE) mid-IR photometry, and a luminous red galaxy (LRG) sample from HSC-SSP. We investigate AGN clustering strength as a function of their luminosity and spectral features across three independent HSC fields totaling $\sim600\,{\rm deg^{2}}$, for $z\in0.6-1.2$ and AGN with $L_{6\mu m}>3\times10^{44}{\rm\,erg\,s^{-1}}$. There are $\sim28,500$ AGN and $\sim1.5$ million LRGs in our primary analysis. We determine the inferred average halo mass for the full AGN sample ($M_h \approx 10^{12.9}h^{-1}M_\odot$), and note that it does not evolve significantly as a function of redshift (over this narrow range) or luminosity. We find that, on average, unobscured AGN ($M_h \approx10^{13.3}h^{-1}M_\odot$) occupy $\sim4.5\times$ more massive halos than obscured AGN ($M_h \approx10^{12.6}h^{-1}M_\odot$), at $5\sigma$ statistical significance {using 1-D uncertainties, and at $3\sigma$ using the full covariance matrix}, suggesting a physical difference between unobscured and obscured AGN, beyond the line-of-sight viewing angle. Furthermore, we find evidence for a halo mass dependence on reddening level within the Type I AGN population, which could support the existence of a previously claimed dust-obscured phase in AGN-host galaxy coevolution. However, we also find that even quite small systematic shifts in the redshift distributions of the AGN sample could plausibly explain current and previously observed differences in $M_{h}$.

\end{abstract}



\section{Introduction} 

Supermassive black holes (SMBH) influence the growth and evolution of the galaxies in which they reside \citep{kormendy_inward_1995, kormendy_coevolution_2013}. Periods of rapid mass accretion onto the SMBH, creating an observable active galactic nucleus \citep[AGN,][]{schmidt_3c_1963}, provide a unique opportunity to study their attributes. Tracing the BH and galaxy coevolution is critical to describing the role of black holes and AGN feedback on galaxy growth \citep{kormendy_coevolution_2013, heckman_coevolution_2014}.

Historically, AGN are classified into two classes, unobscured (Type I) and obscured (Type II). In strict unification models, all AGN are identical but are seen from different angles. Dusty flattened regions on parsec scales, known as the torus, act as a screen at particular inclinations \citep{antonucci_unified_1993, urry_unified_1995, netzer_revisiting_2015}. However, there have been significant results that challenge a strict unified model of AGN structure, showing that the observed obscuration level is correlated with the evolutionary stage of the host galaxy and obscuration effects \citep{sanders_ultraluminous_1988, hickox_clustering_2011, allevato_clustering_2014, ellison_definitive_2019, fawcett_striking_2023}. There have been hints that observed AGN spectral properties are tied to merger history (see \cite{hickox_obscured_2018} for a recent review). Additional studies are needed to probe whether obscured and unobscured AGN are objects along different points in an evolutionary track \citep{hopkins_cosmological_2008, hickox_host_2009, cappelluti_clustering_2012}.

Several observational results have shown that obscured AGN are often found in mergers, suggesting that there is a link with AGN activity  \citep{mihos_triggering_1994, mihos_gasdynamics_1996, blain_dust-obscured_1999, urrutia_evidence_2008, koss_merging_2010, ellison_galaxy_2011, ellison_galaxy_2013, glikman_major_2015, ellison_definitive_2019, goulding_galaxy_2018, secrest_x-ray_2020, ricci_hard_2021}. There is also some evidence that reddening in AGN is significantly impacted by host galaxy properties \citep{goulding_towards_2009, goulding_deep_2012}. Meanwhile, unobscured AGN may be a later stage of the merger evolutionary scenario, once winds and outflows (the potential ``blowout'' phase) have revealed the AGN broad-line region \citep{hopkins_cosmological_2008} and see \cite{almeida_nuclear_2017} for a recent review of AGN obscuration morphology. 

Analyzing these trends in single objects is difficult because the timescales for star formation and mergers are much longer than AGN variability timescales \citep{hickox_black_2014}. However, by measuring the clustering of a statistical sample of galaxies, we can infer the properties of the dark matter (DM) halos in which they reside. Clustering measurements from large scale structure analyses have thus become a preeminent tool to study ensemble properties across samples of AGN and other galaxy populations \citep{osmer_three-dimensional_1981, shaver_clustering_1984, shanks_spatial_1987, iovino_clustering_1988, andreani_evolution_1992, mo_quasar_1993, shanks_qso_1994, croom_qso_1996, la_franca_quasar_1998,  croom_2df_2005, lidz_luminosity_2006, shen_biases_2008, toba_clustering_2017, he_clustering_2018,arita_subaru_2023}. They are particularly useful to study the relationship of galaxies and the halos they are found in, and provide important clues at a population level. 

Studies find that luminous AGN often reside in dark matter halos of typical mass $\log M_h \sim 12.5-13 \,\log h^{-1}M_\odot$ \citep{cappelluti_clustering_2012, timlin_clustering_2018}. These analyses also probe the possible halo mass dependence of observed AGN luminosities, and how it relates to BH accretion properties \citep[e.g.,][]{croom_2df_2005, lidz_luminosity_2006, shen_cross-correlation_2013, mendez_primus_2016, koutoulidis_clustering_2013, krumpe_spatial_2018, timlin_clustering_2018}. 

One way to test the idea that AGN move from a buried (and obscured) growth phase to an unobscured state through blow-out could be to find differences in clustering strength between obscured and unobscured AGN. This would also be a test of the pure orientation model. Simple unification cannot explain the observed differences in clustering amplitude between obscured and unobscured AGN that have already been reported {\citep{hickox_clustering_2011, allevato_clustering_2014, dipompeo_angular_2014, dipompeo_updated_2016, jiang_differences_2016, dipompeo_characteristic_2017, koutoulidis_dependence_2018, powell_swiftbat_2018, petter_host_2023}}. Unfortunately, different studies come to different conclusions about whether obscured or unobscured AGN are found in more massive halos. The results depend on the selection method, luminosities, and redshift range of the samples. On the other hand, many of the studies are based on relatively small areas and only hundreds of AGN \citep[cf.][]{coil_aegis_2009, gilli_spatial_2009, cappelluti_active_2010, allevato_xmm-newton_2011, koutoulidis_clustering_2013,  krumpe_spatial_2018}.

In this work, we implement the AGN selection by Goulding et al. (in-prep.) that aims to represent a more complete sampling of AGN color-space. Using an optical and mid-infrared (MIR) color selection based on unsupervised machine learning classification, we construct unobscured and obscured AGN samples to investigate the clustering strength of these populations. Increasing the source number density will also reduce potential biases in AGN sub-type samples, and allows for more precise measurements of the clustering on all relevant scales. We carry out our analysis using a cross-correlation function between the Hyper Suprime-Cam photometrically selected galaxy sample and our AGN samples, as opposed to the more common AGN autocorrelation. Cross-correlations have the benefit of being less sensitive to systematic uncertainties that are not shared between both samples, and in this case provide a well-understood galaxy sample with which to compare to the AGN. They also have higher S/N for a sparse sample like the AGN. We make use of the individual redshift distribution for each object in our sample, either from photometric or spectroscopic measurements.

This paper is organized as follows. In Section \ref{sec:data}, we summarize the datasets used in this analysis, and the subdivisions of the AGN sample. In Section \ref{sec:methods}, we outline our methodologies for the projected angular correlation function calculation, uncertainty estimation, and parameter fitting and interpretation. We present the results of our autocorrelation of HSC galaxies, the cross-correlation with the full AGN sample, and the cross-correlations with the AGN sub-type samples in Section \ref{sec:results}. We discuss our results in Section \ref{sec:disc}, and conclude in Section \ref{sec:conclu}.

Throughout this analysis, we adopt a $\Lambda$CDM ``Planck 2018''-like cosmology \citep{planck_collaboration_planck_2020}, with $h = H_0/100\,{\rm km\, s^{-1} Mpc^{-1}} = 0.67$, $\Omega_c = 0.27$, $\Omega_b = 0.045$, $n_s = 0.96$, and $\sigma_8 = 0.83$. Quantities expressed with a $\log$ are exclusively $\log_{10}$ values. All magnitudes are in the AB system \citep{oke_secondary_1983}, unless otherwise noted. In the context of galaxy bias and halo mass parametrization, we make use of the \cite{tinker_large-scale_2010} formalism with $\Delta = 200${ (the spherical overdensity radius definition)}. Foreground dust extinction corrections are applied to all magnitudes as supplied in the HSC catalog \citep{aihara_third_2022} based on \cite{schlegel_maps_1998}.

\section{Data} \label{sec:data}

\begin{figure*}
    \centering
    \includegraphics[width = \linewidth]{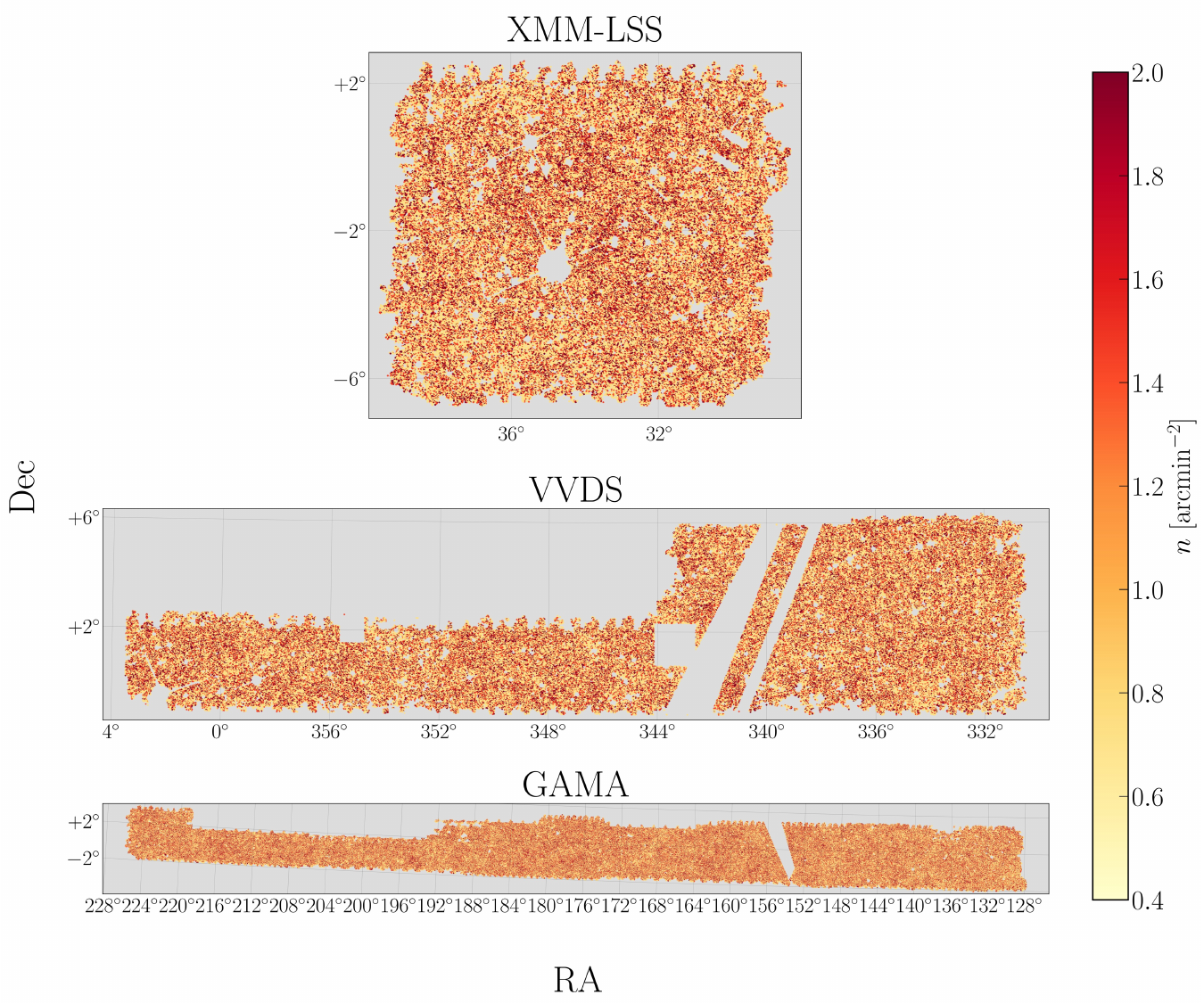}
    \caption{The HSC galaxy sample used in this analysis after isolating the LRG population with an optical color-color cut, split by HSC field. The fields and approximate central R.A., Dec. coordinates are XMM-LSS ($34^\circ , \, -2^\circ$), VVDS ($346^\circ , \, 2^\circ$), and GAMA ($176^\circ , \, 0^\circ$). The bespoke (see \S \ref{sec:masking}) masking strategy, visualized here by stripes that have been removed from the HSC fields, is reflective of areas where there are correlated photometry errors in the WISE data. Other removed areas are primarily because of bright stars and Galactic foreground objects. This figure is produced with the plotting tool \texttt{skymapper} {\citep{melchior_skymapper_2021}}.}
    \label{fig:HSC_fields}
\end{figure*}

\subsection{HSC Survey}

Hyper Suprime-Cam (HSC) is a wide-field prime focus camera mounted on the 8.2\,m Subaru Telescope, located atop Maunakea, Hawai'i \citep{miyazaki_hyper_2018}. The HSC - Subaru Strategic Program (SSP) \citep{aihara_hyper_2018} is designed to make the most of the 1.77\,deg$^2$ field of view by using 330 nights on Subaru to explore the full range of galactic history from the present to $z\sim7$ across three imaging layers, each with specific scientific goals. The Wide, Deep, and UltraDeep surveys consist of different sky coverage and exposure times in the \textit{grizy} wide-band filters, in combination with survey-specific narrow band filters \citep{kawanomoto_hyper_2018}. We refer the reader to the most recent (PDR3) release of HSC-SSP data as described in \cite{aihara_third_2022}. We make use of the 670\,deg$^2$ full-depth full-color Wide imaging within PDR3 to constrain the properties of galaxies positioned across $\sim 2\%$ of the sky. The sensitivity is limited to \textit{i}-band $\lesssim 26$, the point spread function (PSF) $5\sigma$ point source depth. The Wide survey is split between four disjoint fields with varying coverage, two equatorial strips for the spring and autumn, as well as the higher declination field of HECTOMAP and the AEGIS calibration field. 

Galaxy photometric redshifts (photo-$z$'s) are measured from the spectral energy distribution (SED) of each object, following the methods outlined in \cite{tanaka_photometric_2018} and the template-fitting algorithm \texttt{Mizuki} \citep{tanaka_photometric_2015}. This procedure recovers $p(z)$, the redshift probability distribution, for each object. We also have spectroscopic redshifts (spec-$z$'s) for a small fraction ($<2\%$) of objects, but these are not distributed homogeneously over the HSC fields. In HSC PDR3, they combine a wide array of publicly available spectroscopic catalogs to maximize the number of spectroscopic redshifts for observed sources, as detailed in \S4.1 of \cite{aihara_third_2022}. We model the $p(z)$ for spec-$z$'s as a narrow Gaussian distribution, with $\sigma = 0.01 \times (1 + z)$ {(wider than typical spec-$z$ modeling to overcome numerical limitations in our halo model clustering tools)}.

\subsection{Masking} \label{sec:masking}

The HSC PDR3 data have a bright star mask already applied to them, as described in detail by \cite{coupon_bright-star_2018, aihara_third_2022}. They use the Gaia DR2 bright star catalog \citep{gaia_collaboration_gaia_2018} to identify bright sources and remove the affected sky regions from the total survey area. These subtractions are calculated from constraining the extent of stellar halos (the extended PSF of a bright source), ghosts (optical reflections inside the camera that are displayed in the recorded image), blooming (electron spillover into neighboring pixels), and channel-stops (diffraction patterns perpendicular to the blooming) in HSC imaging. This source mask is already applied to our HSC galaxy catalog, and we apply it to the AGN catalog used here.

Details of the source mask constructed based on the WISE imaging data \citep{wright_wide-field_2010, cutri_vizier_2012} can be found in Goulding et al. (in-prep.), we provide a brief summary here. Using the WISE catalog, we select all objects with CCFLAGS=`H {\tt or} D {\tt or} X {\tt or} P'. These catalog objects are those identified from the pipeline to be affected by data issues. We use the distribution of these objects to create a 2D sky density map in order to flag regions of the sky with potentially spurious and/or inaccurately measured sources, removing $\sim 56$\,deg$^2$. {These regions are identified as $\geq 7 \sigma$ overdensities (relative to the average number density of the field) on a map smoothed with a $0.5^{\circ}$ boxcar kernel.} In addition to this flag map, we identify regions of striping within the remaining unmasked objects caused by significantly deeper data, moonlight contamination or additional artefacts. These extremely high-density regions are also excluded as part of the final source mask.

Additionally, we remove point sources at the catalog level that were not included in the bright star mask by identifying and excluding the objects whose \texttt{CModel} magnitudes differed {by  $\leq 0.06$ in \textit{g,r,i}, or \textit{z}} from the PSF magnitude estimate. This difference reflects how well the PSF model for a point source is consistent with the more robust \texttt{CModel} magnitude for an extended object \citep{bosch_hyper_2018}. {Not surprisingly, the excluded objects} follow the stellar locus in color-color space.

\subsection{HSC Galaxy Sample} \label{sec:galaxies}

\begin{table*}
\centering
\caption{Field properties and number of objects in our redshift range ($z\in 0.6-1.2$). }
\begin{tabular}{c|ccc}

\hline \hline  

&GAMA & VVDS&XMM\\
\hline
Area [deg$^2$]  & 397.18  &  100.95 &70.42 \\

$N_{obj}$ (LRGs)  & 1,165,141 (15,444)$^\dagger$ &  303,143 (11,469)$^\dagger$  &204,849 (16,396)$^\dagger$ \\

$N_{obj}$ (AGN)$^*$  & 23,244 (1,981)$^\dagger$ & 5,447 (1,632)$^\dagger$ &  5,457 (1,667)$^\dagger$ \\

$N_{obj}$ (unobscured AGN)$^*$  & 9,197 (1,506)$^\dagger$ & 1,446 (1,074)$^\dagger$  & 1,466 (1,018)$^\dagger$  \\

$N_{obj}$ (reddened AGN)$^*$  & 5,957 (335)$^\dagger$ & 1,150 (392)$^\dagger$ & 1,129 (364)$^\dagger$ \\

$N_{obj}$ (obscured AGN)$^*$  & 7,536 (140)$^\dagger$ & 2,774 (166)$^\dagger$  &2,803 (285)$^\dagger$ \\ 

\hline \hline 

\end{tabular} 

\raggedright
\vspace{0.05in}
\footnotesize
      $^*$ Luminous AGN sample ($L_{6\mu m} > 3\times 10^{44}$ erg s$^{-1}$)

      $^\dagger$ Number of objects with a measured spectroscopic redshift indicated in parentheses.

    \label{tab:field_prop} 
\end{table*}

We use a magnitude-limited galaxy sample ($i<24$) from the HSC-SSP Wide survey, which will be cross correlated with our (HSC-derived) AGN sample. We perform the experiment on the three larger HSC PDR3 equatorial fields with full filter coverage, {XMM-LSS} (hereafter referred to as XMM), VVDS, and GAMA. These fields encompass $2.3 \times 10^7$ galaxies in our analysis redshift range ($z\in 0.6-1.2$, see \S \ref{sec:zbins}), after applying all relevant survey masks (see \S \ref{sec:masking}). The mean number density of galaxies is $\sim 40,000\,{\rm deg^{-2}}$, for a total area of 568.55 deg$^{2}$ (post-masking). Using either spectroscopic or photometric information, we recover the $p(z)$ for each source. 

Prior HSC analyses have selected samples with smaller photo-$z$ uncertainty in order to perform statistical measurements, and found the most effective means of doing so was to use a luminous red galaxy (LRG) sample \citep[e.g., ][]{rau_weak_2023}. Additionally, the strongly clustered LRGs provide an high S/N comparison for cross-correlations. We similarly find it necessary to isolate the LRG population in our galaxy sample to reduce the $p(z)$ uncertainty in our sample, given tests of the galaxy autocorrelation analysis for the complete ($i<24$) HSC galaxy sample. We select the LRG population in the HSC sample with an optical color-color cut ($g-r > 1.2$ and $r-i > 1.0$), and use these objects throughout our clustering analysis cross-correlations. See \S \ref{sec:colorcut} for further details of the LRG selection and preliminary autocorrelation analysis. In total, there are $1.6\times 10^{6}$ LRGs in our analysis, with an average number density of $\sim 2,900$ deg$^{-2}$. Figure \ref{fig:HSC_fields} illustrates the LRG population and its density across the HSC fields in our analysis and Table \ref{tab:field_prop} gives the number of objects in each field.

\subsection{WISE and HSC Selected AGN Sample} \label{sec:AGNsample}

\begin{figure}
    \centering
    \includegraphics[width = \linewidth]{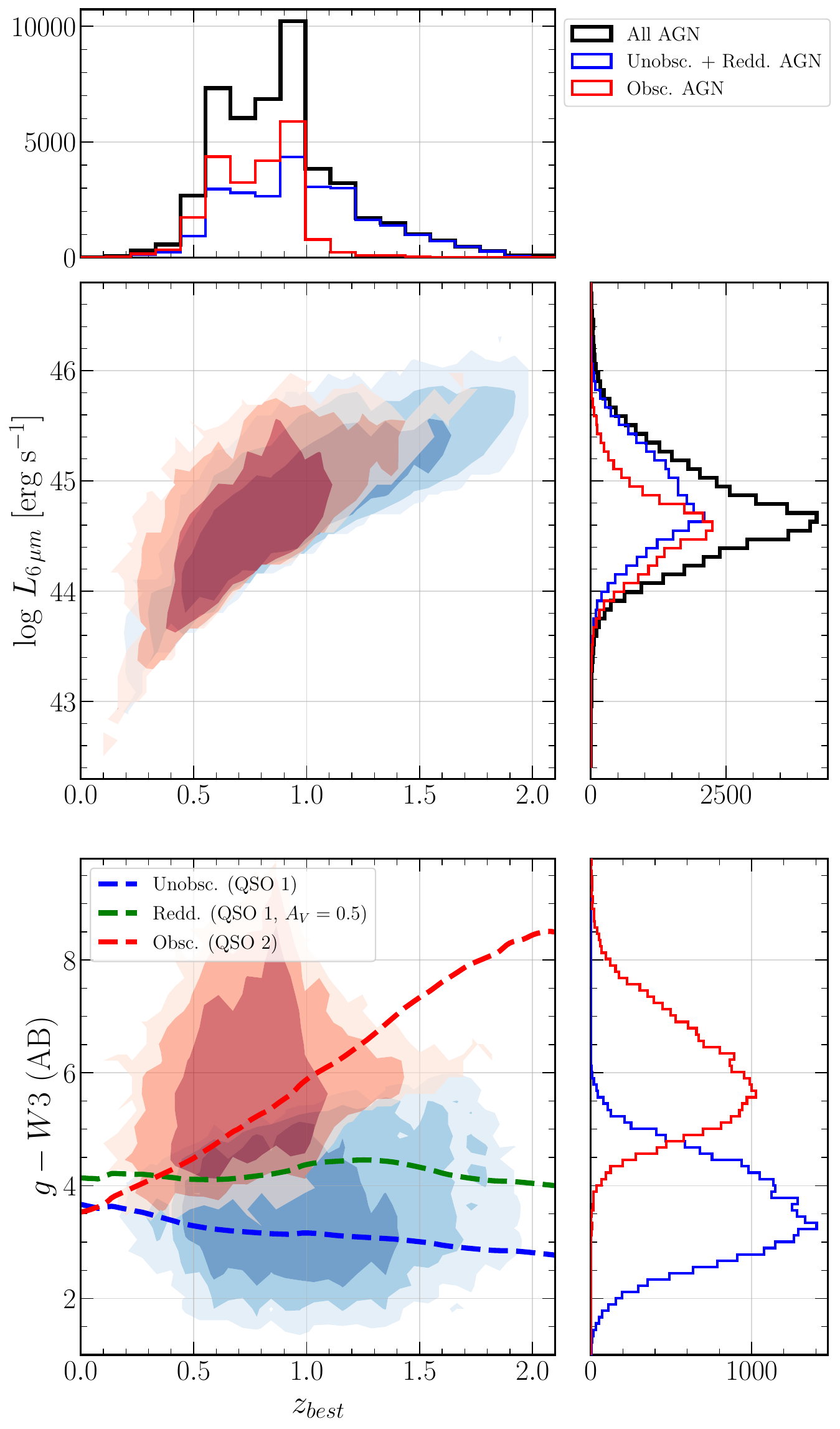}
    \caption{\textit{Top:} The luminosity distribution of the unobscured + reddened (blue contours, objects we infer to have broad emission lines) and obscured (red contours, objects inferred to have narrow emission lines) AGN samples as a function of the fiducial redshift ($z_{best}$) for each object. The contours are set at the (0.25, 0.5, 0.75) quantiles of each distribution. \textit{Bottom:} The $g-W3$ (AB mag) color versus $z_{best}$ distribution for unobscured + reddened and obscured AGN, following the same color scheme and contour quantile levels as in the upper panel. We divide the AGN sample into unobscured, reddened, and obscured classes based on their distribution in this color-redshift space (Goulding et al., in-prep.). We convert the WISE magnitudes in the Vega system to AB with $m_{AB} =m_{Vega} + \Delta m$, where $\Delta m (W3) = 5.174$ \citep{jarrett_spitzer-wise_2011}. The dashed lines are the $g-W3$ color redshift evolution for three AGN templates from \cite{polletta_spectral_2007}. We use their unobscured quasar template (QSO1, blue line), their unobscured quasar with a \cite{draine_interstellar_2003} extinction law applied with $A_V = 0.5, R_V = 3.1$ to create a reddened quasar template (green line), and their obscured quasar template (QSO2, red line). Histograms show the individual distributions for a given quantity, where colors indicate the AGN sample specified.}
    \label{fig:color_Lz}
\end{figure}

\begin{figure}
    \centering
    \includegraphics[width = 0.94\linewidth]{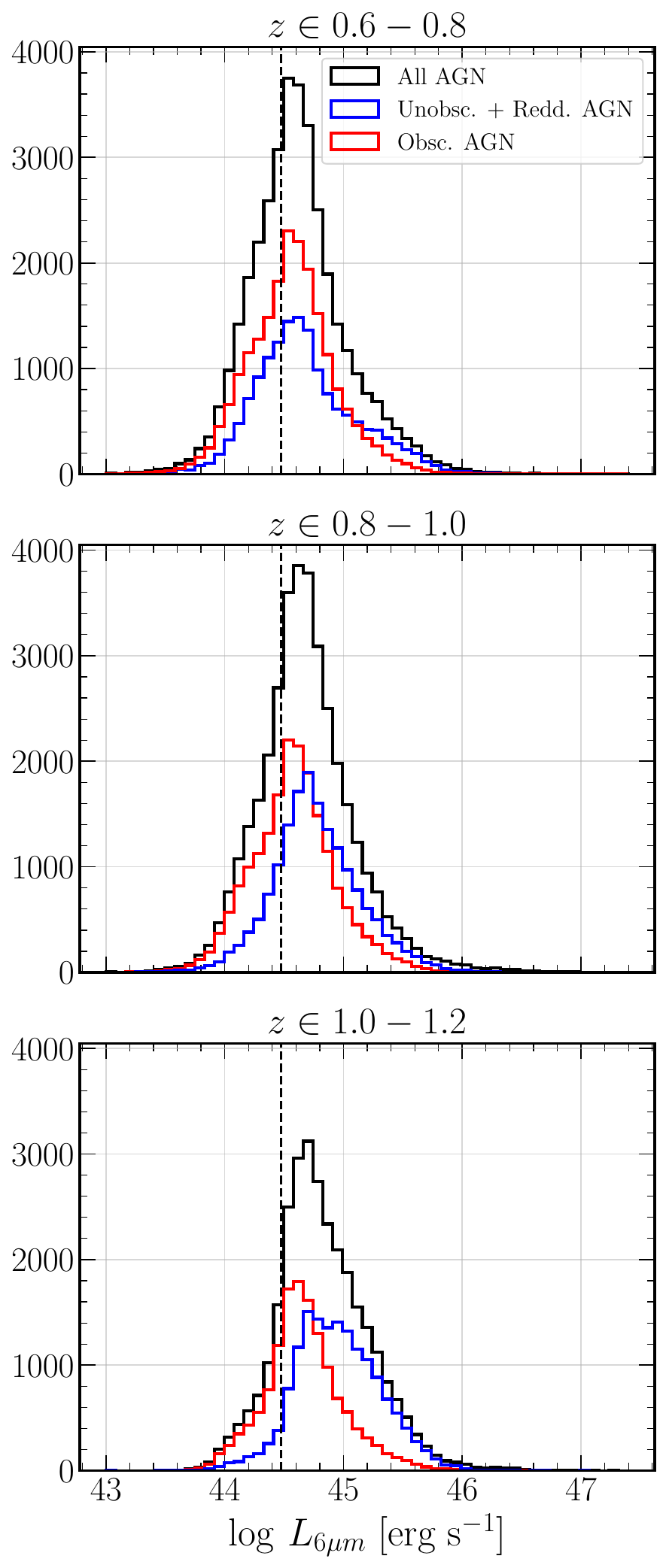}
    \caption{$L_{6 \mu m}$ luminosity distributions for the specified redshift bins for this analysis combining all considered HSC fields. The vertical dotted line indicates a $L_{6\mu m}$ lower limit for this analysis at $3\times 10^{44}\,{\rm erg\,s^{-1}}$. The different colors indicate the different AGN subsamples.}
    \label{fig:L6dists}
\end{figure}

\begin{figure}
    \centering
    \includegraphics[width = 0.94\linewidth]{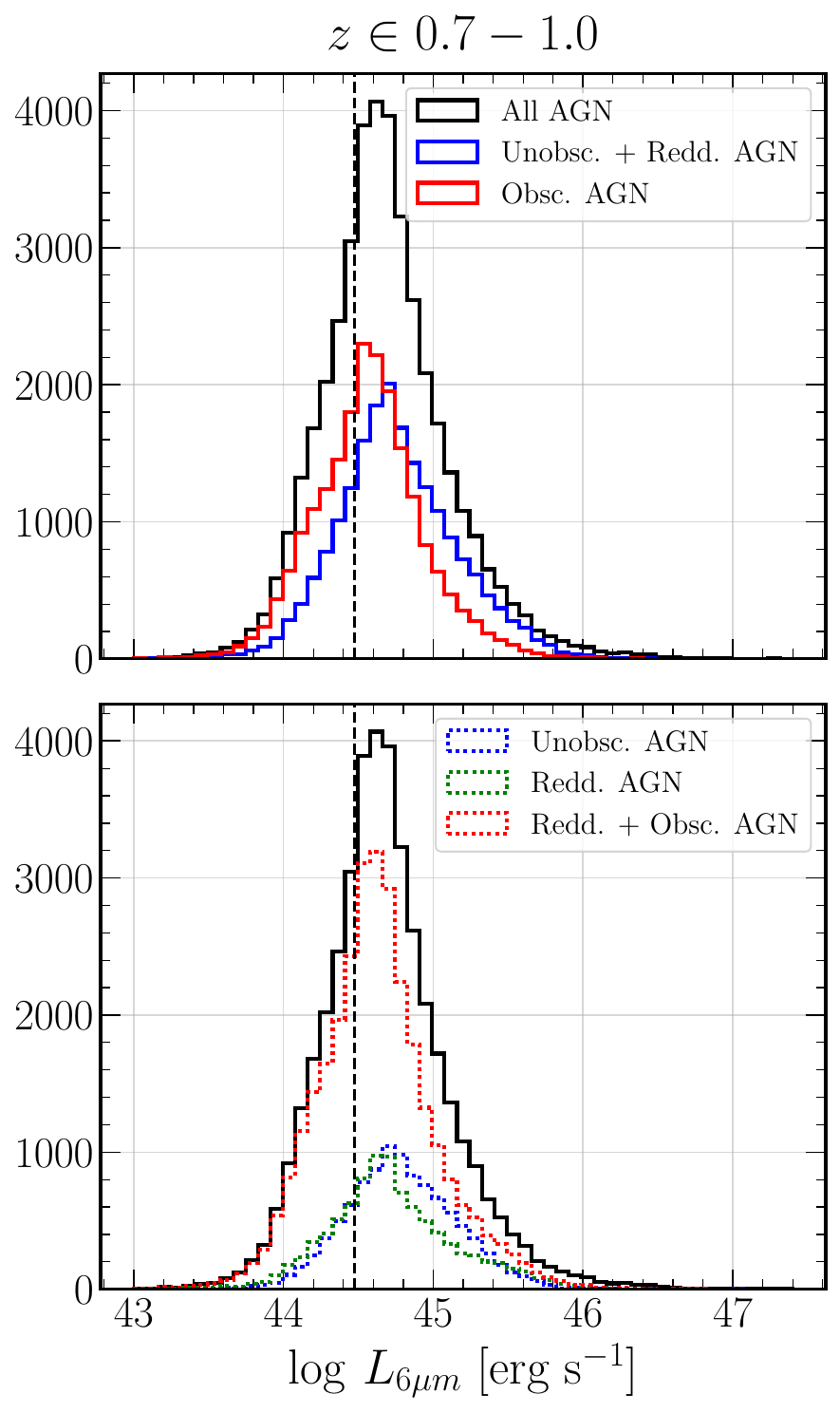}
    \caption{$L_{6 \mu m}$ luminosity distributions for the wide redshift bin for this analysis combining all considered HSC fields. The vertical dotted line indicates $L_{6\mu m}$ lower limit for this analysis at $3\times 10^{44}\,{\rm erg\,s^{-1}}$. The different colors indicate the different AGN subsamples that are considered in the wide redshift bin analysis.}
    \label{fig:L6dists_W}
\end{figure}

We present an AGN sample that combines optical and mid-IR photometry from HSC-SSP and the Wide-field Infrared Survey Explorer (WISE) \citep{wright_wide-field_2010}. The details of this method are explained in Goulding et al. (in prep.), and we give a high-level overview here. Using a maximum likelihood estimator, HSC \textit{grizy} photometry is matched to sources detected with S/N$>5$ in their W1 photometry in the allWISE \citep{wright_allwise_2019} and unWISE \citep{unwise_team_unwise_2021} catalogs \citep{mainzer_preliminary_2011, cutri_vizier_2021, schlafly_unwise_2019}. After combining the source catalogs, we additionally require sources to have S/N $>4,3,3$ in their $g$, $W2$ and $W3$ photometry.

Utilizing an unsupervised dimensionality reduction technique, the Uniform Manifold Approximation \& Projection (UMAP) algorithm \citep{mcinnes_umap_2018}, we distill the multi-dimensional color, magnitude, and source size space down to a single interpretable two-dimensional manifold. UMAP incorporates the input multi-dimensional photometric information from HSC and WISE to construct a neural network that identifies objects with similar properties, grouping them together, while simultaneously placing objects with dissimilar properties far apart. The result is a simple 2D space in which we can identify the region occupied by AGN. Inputting a test sample with known labels after the training process has been completed, but providing no prior knowledge of the intrinsic source properties to UMAP, Goulding et al. (in-prep.) show that UMAP segregates known stars from galaxies, and known obscured + unobscured AGN/quasars from inactive galaxies. The final UMAP manifold contains 16 distinct clusters, three of which are dominated by AGN. The AGN-nature of these clusters are further validated via spectroscopic follow-up by \cite{hviding_spectroscopic_2024}, who find the UMAP AGN classification matches the spectral categorization with emission line fitting and continuum ratios indicative of AGN activity. {These UMAP AGN sub-type classifications are not part of the principal AGN identification  detailed in \S \ref{sec:lum_bins}. We use this step to solely identify the AGN.}

Goulding et al. (in-prep.) further provide photometric redshift distributions and spectroscopic redshifts (where available) for all of the UMAP-classified AGN. These photometric redshifts are determined by utilizing the full \textit{g} through \textit{W3} photometry to train an augmented Random Forest algorithm, which they show significantly outperforms the {\texttt{Mizuki}} photo-$z$'s that are primarily designed for inactive galaxies. These Random Forest-based photometric redshifts perform equally well on both Type I and Type II AGN alike out to $z \sim 3$ with an average precision of $\delta z$/(1+$z$)$\sim 0.02$ and 0.03, respectively.

The total redshift distribution for the AGN samples from photometric and spectroscopic measurements are illustrated in the top panel of Figure \ref{fig:color_Lz}. We account for the each object's full $p(z)$ in the clustering measurements, as described in \S \ref{sec:weightedmethod}.

\subsubsection{AGN Subsamples}\label{sec:lum_bins}
Accurate AGN classification is paramount to appropriately estimate the objects' properties. Previously, AGN sub-types have been split based on an optical-MIR color, such as $r-$W2 \citep[c.f.][]{petter_host_2023}. However, a more robust classification scheme should consider the SED evolution with redshift, and empirically derive the minima in the AGN color-redshift space to sort the sub-types{, which we now proceed to do.}

From the subset of the UMAP-classified AGN with spectroscopy available from the Sloan Digital Sky survey, Goulding et al. (in-prep.) train a K-Nearest Neighbor (KNN) algorithm to probabilistically label these AGN as unobscured (unobsc.), reddened (redd.), or obscured (obsc.), based on their position in $g-W3$ vs. redshift space (the output of these distributions are shown in Figure \ref{fig:color_Lz}). The unobscured and reddened training samples are spectroscopically characterized by having broad emission lines, but the reddened AGN have significant dust obscuration and thus a red continuum in the rest-frame optical. We will typically combine these in our analysis as an unobscured + reddened sample. Meanwhile, the obscured AGN training sample is made up of objects that are both heavily dust obscured and have only narrow emission lines. This process recovers $\sim48,000$ AGN in our redshift range (i.e. at least $3\%$ of the $p(z)$ is in $z\in 0.6-1.2$), across the three HSC fields used here. Of these, $\sim15,000$ objects are unobscured AGN, $\sim 11,000$ are reddened objects, $\sim21,000$ are obscured AGN, and $<1\%$ of objects do not have a spectral type assigned. Probabilistic estimates of AGN/galaxy classifications using machine learning methodologies that encompass a wide range of source properties inherently come with caveats, but on the whole are deemed to be more complete and reliable than standard singular or 2-D demarcations. Hence, we choose to exploit the available data to its reasonable limits to classify our AGN.

The $g-W3$ distribution of the unobscured + reddened and obscured objects in our analysis as a function of redshift is shown in the bottom panels of Figure \ref{fig:color_Lz}. The redshift evolution for the $g-W3$ color of SWIRE library templates \citep{polletta_spectral_2007} for three types of AGN are shown in dashed lines in the bottom panel of Figure \ref{fig:color_Lz}. We will investigate the clustering amplitude of the sets of unobscured + reddened and obscured AGN, as well as the difference between unobscured and reddened AGN. 

We note that these color cuts are a function of redshift, given the minima of the KNN probability density distribution, and are not like previously defined selection functions for AGN types using only optical and MIR colors \citep[eg.,][]{stern_midinfrared_2005, hickox_host_2009, stern_mid-infrared_2012, assef_mid-infrared_2013, hviding_new_2022}. These new classifications recover a significant fraction of spectroscopically-confirmed $W1 - W2 < 0.8$ AGN (below the \cite{stern_mid-infrared_2012} limit), generating a wider (and more complete) sampling of unobscured + reddened and obscured AGN color-space for our analysis \citep[AGN spectra from SDSS DR16, ][]{ahumada_16th_2020}. From follow-up spectra of 178 objects in our sample with $i<22.5$, \cite{hviding_spectroscopic_2024} confirm the classification for $85\%$ of the randomly selected unobscured + reddened AGN, and similarly $65\%$ of the obscured AGN sample, finding a contamination rate of $3\%$ and $15\%$, respectively, and bolstering our confidence in the classification.

We present the subsample redshift and luminosity distributions in Figure \ref{fig:color_Lz}. We note the excess of unobscured + reddened objects at high redshift, which are driving the excess of high luminosity AGN in the $L_{6\mu m}$ plot (see middle left panel of Figure \ref{fig:color_Lz}). We obtain rest-frame $L_{6 \mu m}$ luminosity measurements for each AGN in the sample via the standard power law fitting to the MIR photometry from WISE. In order to compare the clustering across AGN sub-types, we divide the analysis into a series of redshift bins (\S \ref{sec:zbins}). Their luminosity distributions are illustrated in Figures \ref{fig:L6dists} and \ref{fig:L6dists_W}. We establish a lower limit for the AGN luminosity at $L_{6 \mu m } > 3 \times 10^{44} \, {\rm erg\,s^{-1}}$, close to the peak of the distribution, where we are confident that we are complete out to $z\sim1.2$ (see Figure \ref{fig:color_Lz}). This has the added benefit of allowing an analysis comparing AGN sub-types where their luminosity distribution is more consistent, above our threshold. This results in a total of $34,144$ luminous AGN in our redshift range that we will use in our analysis. The number of luminous AGN (and their spectral classes) in each field are shown in Table \ref{tab:field_prop}, including the number of objects for which we have spectroscopic redshifts. 

Additionally, we define a low and high $L_{6 \mu m}$ range to investigate if there is any evidence of an inferred halo mass trend with luminosity. These additional cuts are set at $ 3 \times 10^{44} < L_{6 \mu m} < 10^{45}\,{\rm erg\, s^{-1}}$ for the low $L_{6\mu m}$ sample, and $L_{6 \mu m} > 10^{45}\,{\rm erg\, s^{-1}}$ for the high $L_{6\mu m}$ sample.

\section{Methodology} \label{sec:methods}
In the following section, we describe our methods to measure the clustering statistic and compare with dark matter halo models to ascertain galaxy bias and halo mass estimates. 

\subsection{2-Point Statistics to Assess Galaxy Clustering}

\subsubsection{Angular Correlation Function}
The two-point angular clustering statistic is defined as the excess probability of a pair of objects being separated by an angle $\theta$ above a Poisson (random) distribution \citep{peebles_statistical_1973}. Spatial correlation functions are a critical tool with which to analyze the clustering properties of galaxies on a wide range of angular scales. With only photometric redshift information in hand for every object in our sample, we limit this analysis to a projected clustering statistic, rather than one in three dimensions. In order to reduce shot noise bias, we employ the \cite{landy_bias_1993} estimator of the angular two-point function: 

 \begin{equation}\label{eq:LS}
    \omega(\theta) = \frac{DD(\theta) - 2 DR(\theta) + RR(\theta)}{RR(\theta)},
\end{equation}
where we further define the data-data, data-random and random-random pair counting operations $DD$, $DR$, $RR$ as:
\begin{equation}
    XY(\theta) = \frac{X Y^\prime(\theta)}{N_{X} N_{Y}}
\end{equation}
to ensure proper normalization given the amount of objects $N$ considered in the operation. While $D$ is drawn from the observational catalogs, $R$ is synthesized as a set of random points. We generate these randoms such that they match the survey footprint and have all area masks applied. This is critical in order to ensure that any excess clustering from the data is measured in the same survey geometry as the random points that establish the comparison. We will be using a weighted pair counting and binning statistic in this analysis, as detailed in \S \ref{sec:weightedmethod}.

We follow \cite{awan_angular_2020} in formalizing the pair counting operation as
\begin{equation}
X Y^\prime(\theta) = \sum_i^{N_X} \sum_{j \neq i}^{N_Y} \bar{\Theta}_{ij,k}
\end{equation}
where 
\begin{equation}
    \bar{\Theta}_{ij,k} = \Theta(\theta_{ij} - \theta_{{\rm min}, k} ) [1 - \Theta(\theta_{ij} - \theta_{{\rm max}, k} )],
\end{equation}
and 
\begin{equation}
    \Theta(x) =
    \begin{cases}
        0 ~ {\rm,~if} ~ x < 0 \\
        1 ~ {\rm,~if} ~ x \geq 0 \\
    \end{cases}.
\end{equation}
The Heaviside step function details the binning operation that counts the number of objects at an angular distance from another object. $\theta_{ij}$ is the separation between galaxy $i$ and $j$ in the sample of $N_X$ and $N_Y$ galaxies. The per-bin counting operator $\bar{\Theta}_{ij,k}$ counts the number of galaxy pairs at separation  $\theta_{{\rm min}, k} \leq \theta_{ij} < \theta_{{\rm max}, k}$ for the $k$th bin. This formalism is equally applicable to the auto- or cross-correlation of galaxy catalogs. We use 24 spatial bins for the calculation, logarithmically spaced from $s = 0.01 \, h^{-1} {\rm Mpc}$ to $s =100 \, h^{-1} {\rm Mpc}$. We convert these projected scales to angular bins with a standard angular diameter distance conversion (from the comoving distance), with the median of the sample's $dN/dz$ as the fiducial $z$. 

\subsubsection{Uncertainty Estimation}
We use a standard jackknife procedure to estimate the total statistical and systematic error in a given per-field analysis. Splitting each of the three fields into 25 equal-area regions, we calculate the clustering signal from a set of 24 regions, removing one region per iteration until all 25 have been removed once. This removes approximately $4\%$ of the data in each calculation. The per-bin error {(that will be used as the $1\sigma$ error bars in each of the measured correlation functions)} is then calculated from the square root of the diagonal of the covariance matrix as defined in \cite{norberg_statistical_2009}:
\begin{equation}
    C_{jk} (x_i, x_j) = \frac{N-1}{N} \sum^N_{k = 1} (x^k_i - \bar{x}_i)  (x^k_j - \bar{x}_j),
\end{equation}
where $x_i$ is the $i$th measure of the statistic, i.e. $\omega(\theta)$ , out of a total of $N =25$ measurements, and 
\begin{equation}
    \bar{x}_i = \sum^N_{k = 1} x^k_i / N .
\end{equation}

{Previous analyses have adopted two different approaches to quantifying the uncertainties, \cite{hickox_clustering_2011, allevato_xmm-newton_2011, allevato_clustering_2014, laurent_clustering_2017, krumpe_spatial_2018} use the full covariance matrix while \cite{koutoulidis_clustering_2013, koutoulidis_dependence_2018, dipompeo_angular_2014, dipompeo_updated_2016, dipompeo_characteristic_2017} use only the square root of the diagonal terms of $C_{jk}$ as the 1-D uncertainty. The most recent analysis of \cite{petter_host_2023} that we compare with here uses the diagonal terms. For ease of comparison with prior work, we adopt the diagonal treatment as our default approach. We also evaluate and report the resulting best fit galaxy bias and inferred halo mass via least-squares minimization of the data--model residual with the full $C_{jk}$ matrix. In doing so, we capture the contribution of the real off-diagonal bin-to-bin correlations from the jackknife, and more accurately capture the precision of our measurements. In presenting the results from the full covariance matrix treatments, we show that our conclusions do not qualitatively change when either approach is implemented (see \S \ref{sec:app_covmat}).} Additionally, the systematic uncertainty as a function of field-to-field variability is constrained by comparing the different fitted bias ($b$) and halo mass ($M_h$) values from each field (defined in full in the following subsection). We combine the bias and halo mass distribution from all fields to estimate the average value and uncertainty from the complete sample via an inverse variance weighted mean.

\subsection{Clustering Interpretation}\label{sec:model_def}
We use a halo model to infer the clustering properties from the measured excess probability $\omega(\theta)$. The relationship between galaxy clustering and dark matter halo clustering is formalized with the multiplicative ratio term of the galaxy bias, where we are working within the Eulerian framework of peak background split theory \citep{sheth_large_1999}. The bias term $b$ describes the excess clustering signal in galaxies relative to a dark matter halo clustering model. We use the measured redshift distribution from our sample to model the dark matter halo clustering signal, which we parameterize with Limber's Equation \citep{limber_analysis_1953, groth_statistical_1977,  peacock_power_1991, eisenstein_correlations_2001}: 

\begin{align} \label{eq:limber}
\begin{split}
    \omega(\theta) = \pi \int_{z = 0} ^{\infty} \int_{k = 0} ^{\infty} \frac{\Delta^2 (k,z)}{ k^2} J_0[k\, \theta \, \chi(z)] 
    \\
    \times \left(\frac{dN}{dz}\right)_1 \left(\frac{dN}{dz}\right)_2 \left(\frac{dz}{d \chi}\right) dk\, dz , 
\end{split}
\end{align}

where we use $\Delta^2 (k,z) = \frac{k^3}{2\pi^2} P_{HF}(k,z)$, $P_{HF}(k,z)$ being the dark matter power spectrum from linear to nonlinear scales from \texttt{halofit} as implemented in the Core Cosmology Library (CCL) \citep{takahashi_revising_2012, chisari_core_2019}. $J_0$ is the zeroth-order Bessel function, $\chi(z)$ is comoving distance in units of $h^{-1}$ Mpc. $dz/d\chi$ is defined as $H(z)/c = (H_0/c) [\Omega_m (1+z)^3 + \Omega_{\Lambda}]^{1/2}$ for a flat cosmology in $h$\,Mpc$^{-1}$.

This parametrization takes a model for the three-dimensional clustering of the halos and the $dN/dz$ from our particular dataset, properly accounting for the expected halo clustering over the redshift range being probed. This synthesis of the individual galaxies' redshift probability distributions $p(z)$ is usually constructed via a simple sum of the individual source $p(z)$,
\begin{equation}\label{eq:dNdz}
    \frac{dN}{dz} = \sum_i p_i(z).
\end{equation}

As detailed in \S \ref{sec:weightedmethod}, we use Equation \eqref{eq:wdNdz} to perform a weighted sum that considers an object's full $p(z)$. The benefit of this weighted method is that it includes the probability of an object being both within and outside a defined redshift bin, rather than standard tomographic approaches that include objects whose fiducial redshift is in the bin. We find that the results from this method are broadly consistent with the standard tomographic method, but opt for our weighted method in order to reduce the possible effects to systematic biases inherent in photometric redshift fitting.

We fit the measured clustering with the forward model via a linear least squares fit with the simple multiplicative scalar value $b^2$, using a Monte Carlo (MC) sampling estimator. For our cross-correlation, the fitted multiplicative value is $b_G b_A$, i.e. the product of the bias from each of the two datasets (galaxies and AGN) being correlated. We calculate the galaxy bias from the autocorrelation, as well as the cross-correlation between the galaxies and AGN samples. In this analysis,  we are able to divide the cross bias term by the recovered galaxy bias $b_G$ and isolate the AGN bias $b_A$. We judge the goodness of fit via a $\chi^2$ test, recording the reduced $\chi^2$ statistic, $\chi^2_\nu = \chi^2/n_{d.o.f.}$, for the total degrees of freedom ($X$ data points being fit - 1 fitting parameter, $b$ or $M_h$). Using the combined non-linear and linear dark matter halo model defined in \texttt{halofit}, we establish our fitting range to include smaller scales than what could be constrained with the linear model alone \citep{zehavi_departures_2004, takahashi_revising_2012}. We do not model the non-linear one-halo term in this work, as that would require a halo occupation distribution (HOD) treatment \citep[e.g.,][]{berlind_halo_2002}, and does not contribute to the linear bias estimate. We thus only fit over the scales $\gtrsim 3\arcmin$ in the measured correlation function, avoiding points on smaller scales in which the more complex HOD modeling would be necessary. This angular extent is chosen such that we fit for $s > 1\,h^{-1}\,$Mpc in all redshift bins. Given these parameters and our 24 logarithmically-spaced angular bins, we have a total of 11 data points from each correlation function to fit over, giving us 10 degrees of freedom (d.o.f.).  We use the implementation of Limber's equation in the LSST Dark Energy Science Collaboration (DESC) Core Cosmology Library (CCL)\footnote{\url{https://github.com/LSSTDESC/CCL.}} to include the measured $dN/dz$ in the \texttt{halofit} model \citep{chisari_core_2019}. This method produces the standard bias analysis in the (mostly) linear regime, and allows for an extension to estimate the mass of the average halo leading to the measured bias. 

With this model and fitting algorithm, we infer the average halo mass of our samples, $M_h$, from the measured linear clustering bias. We use the \cite{tinker_large-scale_2010} parametrization of the halo mass function to infer the halo masses traced by the measured bias values, as do \cite{laurent_clustering_2017}. Following \cite{tegmark_time_1998}, we can use the analytical form of $b(z,M)$ such that we may replace $P_{HF}(k,z)$ in Equation \eqref{eq:limber} with $b^2(z,M)\, P_{HF}(k,z)$ in an autocorrelation, fitting for the parameter $M$ using a Monte Carlo analysis. From this estimate of the galaxy bias and halo mass, we may then calculate the AGN halo mass by replacing the $P_{HF}(k,z)$ in Equation \eqref{eq:limber} with $b_G \, b_A(z,M) \, P_{HF}(k,z)$, where $b_G$ is known from the autocorrelation. While more uncertain than the bias measurement due to the several assumptions made for the halo mass function and halo occupation distribution, a mass inference will isolate the halo properties from the implicit redshift evolution in the recovered bias. Moreover, halo mass comparisons across analyses is complicated by subtle differences in the precise formalism of the bias-halo mass translation used, and can introduce significant systematic shifts.

{Following \cite{laurent_clustering_2017}, we match} the measured linear bias with an MC-derived halo mass best fit {to estimate} $M_{h,\,min}$, i.e. a model for a strict lower limit in $M_h$. For every measured bias value in our chain, we calculate the $M_{h,\,min}$ by solving for this value in
\begin{equation}
    b_1(z,M_{h, min}) = \frac{\int^{\infty}_{M_{h, min}}  \frac{dn}{dM} \langle N(M) \rangle b(z,M) dM }{\int^{\infty}_{M_{h, min}} \frac{dn}{dM} \langle N(M) \rangle dM} , 
\end{equation}
where $b_1$ is the measured clustering bias, $dn/dM$ is the halo mass function, the effective bias function $b(z,M)$ is as defined in \cite{tinker_large-scale_2010}, and $\langle N(M) \rangle$ is the average halo occupation, which we assume is 1 for our sample (i.e. every AGN is in the central galaxy of its halo). 

This defines the lower bound of the mass range of halos, which we can then use to estimate the average halo mass, $\langle M_h \rangle$, folding in the halo mass and redshift distribution of our sample:
\begin{equation}
    \langle M_{h} \rangle = \frac{\int^{\infty}_{M_{h,\,min}} \int_{z} M \frac{dn}{dM} \frac{dN}{dz} \langle N(M) \rangle\, dz \,dM  }{\int^{\infty}_{M_{h,\,min}} \int_{z} \frac{dn}{dM} \frac{dN}{dz} \langle N(M) \rangle\, dz \,dM } .
\end{equation}

This average halo mass estimate is more representative of the complete halo mass distribution that leads to the measured galaxy bias. Inferring the average host halo mass through this weighted average properly accounts for the cosmological distribution of halo masses in the Universe, as opposed to assuming all AGN are hosted in halos of a single representative mass $M_{\rm eff}$. {We discuss intercomparison of different results in light of possible choices for the $b - M_h$ connection formalism in \S \ref{sec:disscussion_1}.}%{We also find that the $M_{\rm eff}$ approach to the halo mass function consistently underestimates the inferred $\langle M_{h} \rangle$ value.}

\subsection{Redshift Bins} \label{sec:zbins}
Following \citep{nicola_tomographic_2020, rau_weak_2023}, we subdivide our total sample into broad redshift bins where we evaluate the properties of the clustering as a function of cosmic time. How these bins are constructed becomes particularly relevant in photometric redshift surveys where the $p(z)$ is much broader than for a spectroscopic measurement. 

Our finalized redshift bins are selected such that we have significant overlap in redshift between the HSC Galaxy sample and the AGN, at $z > 0.5$. The median uncertainty of the LRG photo-$z$ estimates, given their $p(z)$, is $\sigma_z \sim 0.1$. This uncertainty leads us to choose a bin width of $\Delta z = 0.2$ in order to capture the full $1\sigma$ distribution of an object in the center of our bin. We define three narrow redshift bins over which to conduct the complete AGN sample analysis. The first bin is delimited at $z\in0.6 -0.8$, the second bin is $z\in0.8 - 1.0$, and the third is defined for $z\in1.0 -1.2$. We construct an additional wider bin with which to investigate the clustering across AGN sub-types, defined for $z \in 0.7 - 1.0$.

\section{Results} \label{sec:results}

\begin{figure*}
    \centering
    \includegraphics[width = 0.99\linewidth]{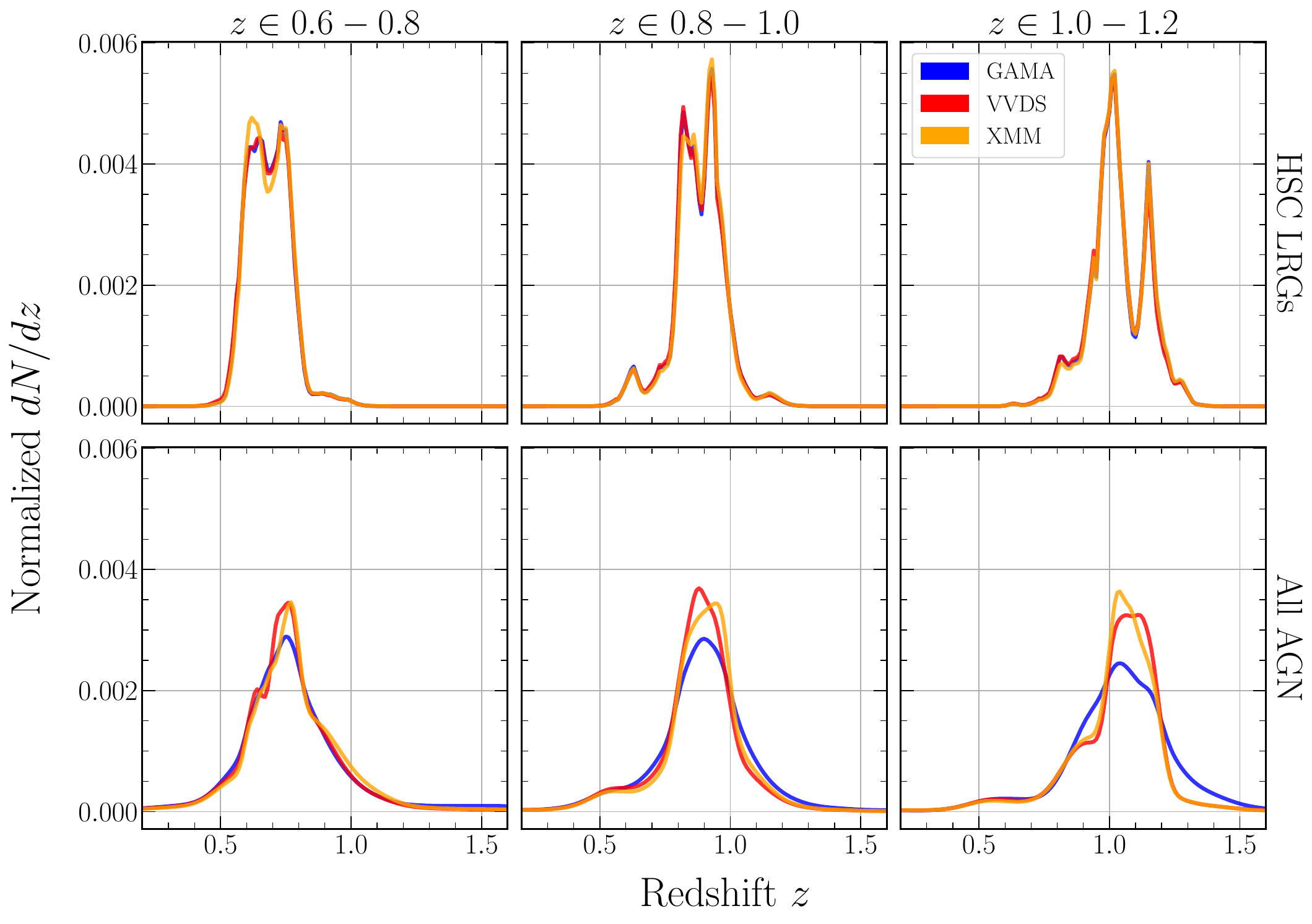}
    \caption{Normalized $dN/dz$ measured from the LRG (top) and full AGN samples ({bottom}) for auto- and cross-correlations in the three narrow redshift bins, across different HSC fields. These are constructed following the procedure outlined in \S \ref{sec:weightedmethod}.}
    \label{fig:autodNdz}
\end{figure*}

\begin{figure*}
    \centering
    \includegraphics[width = 0.99\linewidth]{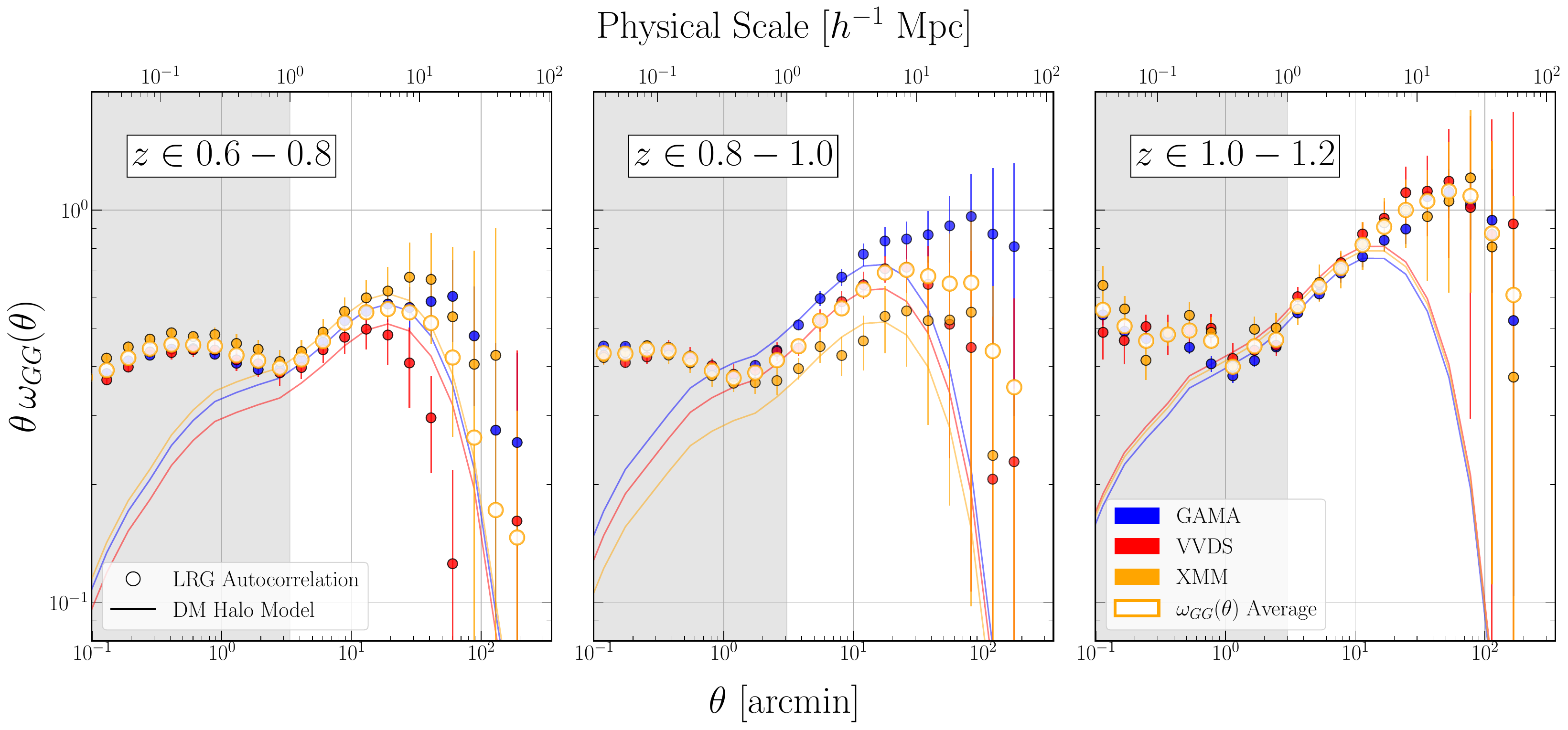}
    \caption{The measured HSC LRG projected angular autocorrelation in our three HSC fields across three redshift bins in $\theta \omega(\theta)$. We use this scaling by a power of $\theta$ to reduce the dynamic range of the plot and more clearly see the differences in the measured points. The open symbols are the per-bin averaged correlation function across fields. {The $1\sigma$ uncertainties are drawn from the square root of the diagonal of the jackknife covariance matrix for the sample.} The solid lines represent the fitted  \texttt{halofit} DM model to each field. We fit points for physical scales $s > 1\, h^{-1}{\rm Mpc}$, while the grey region contains the interior points excluded from the fit. Note that the model does not include the 1-halo term, which is why the data rise significantly above the model on small scales.}
    \label{fig:autocorr}
\end{figure*}

\begin{figure*}
    \centering
    \includegraphics[width = 0.99\linewidth]{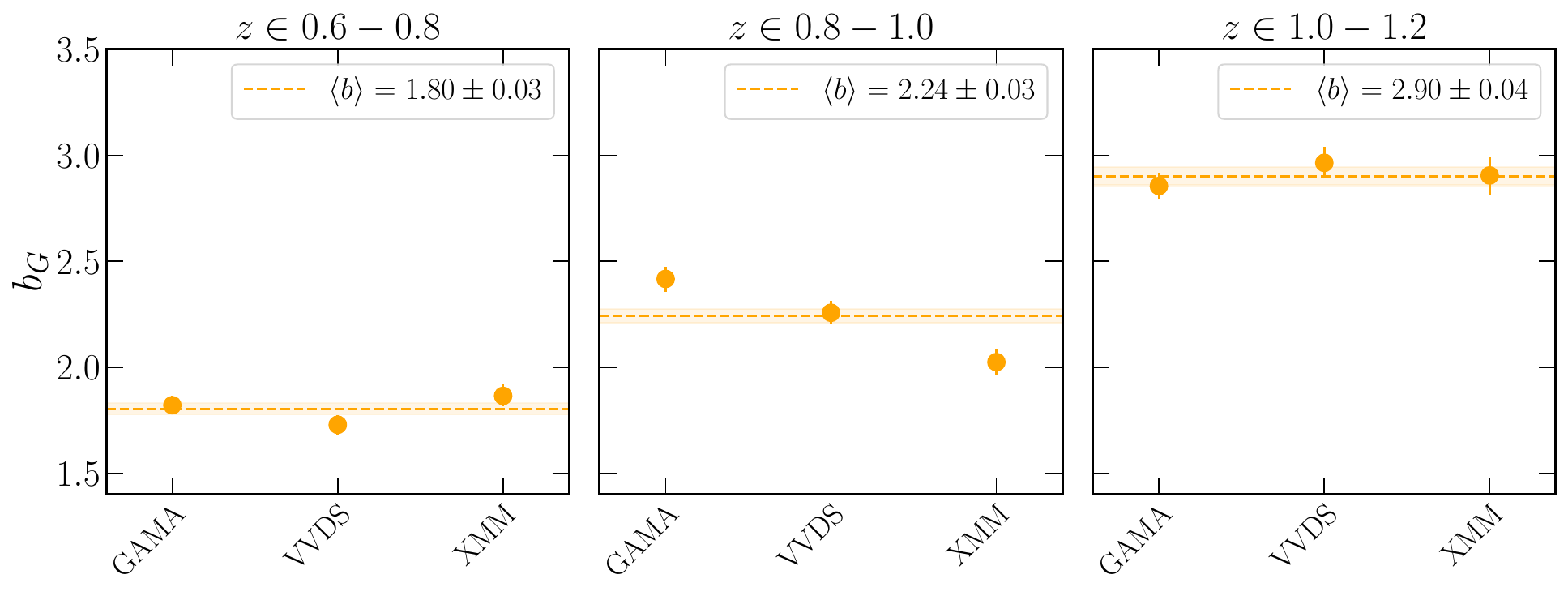}
    \caption{Recovered galaxy bias from the autocorrelation of HSC-selected LRGs, in three redshift bins, for each analysis field. The inverse variance weighted mean and $1\sigma$ uncertainty are plotted with the dashed line and shaded region.}
    \label{fig:auto_summ}
\end{figure*}

\begin{figure*}
    \centering
    \includegraphics[width = 0.99\linewidth]{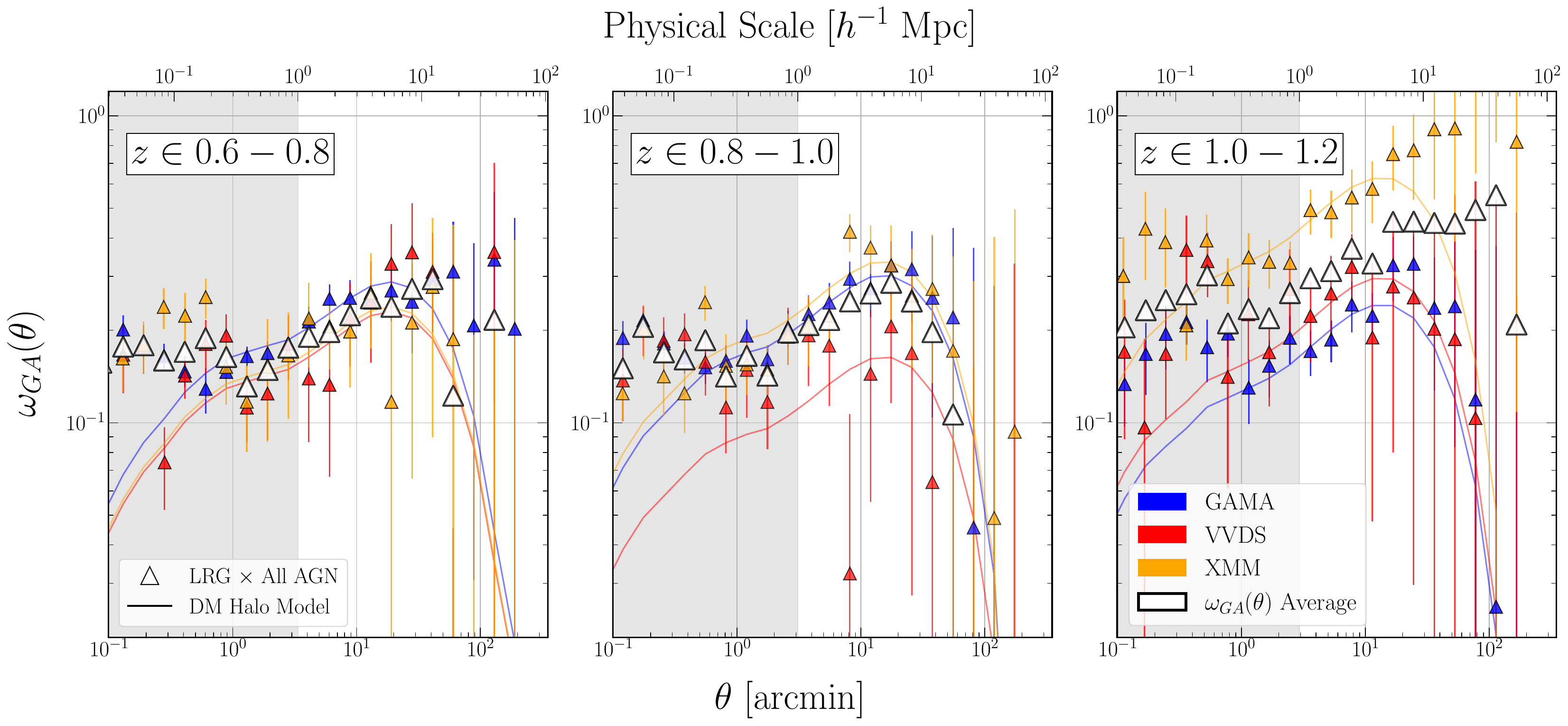}
    \caption{Cross-correlation between HSC LRGs and all AGN measured, and fitted with CCL in different fields across redshift bins, shown in $\theta \omega(\theta)$ to reduce the plotted dynamic range. The AGN sample is limited to $L_{6\mu m} > 3\times 10^{44} \, {\rm erg\, s^{-1}}$. The open symbols are the per-bin averaged $\omega_{GA}(\theta)$ across fields. {The $1\sigma$ uncertainties are drawn from the square root of the diagonal of the jackknife covariance matrix for the AGN sample.}The solid lines represent the fitted \texttt{halofit} DM model to each field. The points considered in this analysis are those past the grey shaded region, at $s > 1 \,h^{-1} {\rm Mpc} \, (\theta \gtrsim 3\arcmin)$. }
    \label{fig:crosscorr_all}
\end{figure*}

\begin{figure*}
    \centering
    \includegraphics[width = 0.99\linewidth]{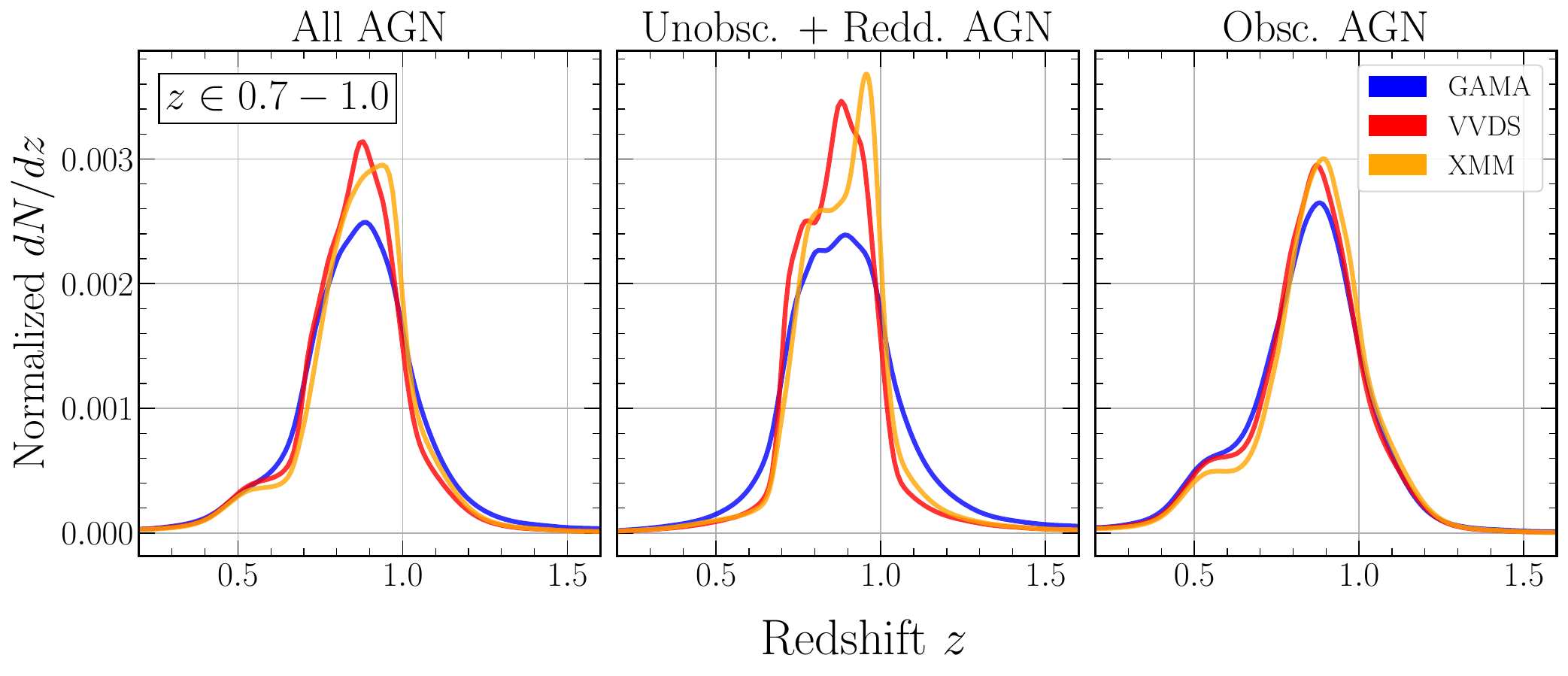}
    \caption{Normalized $dN/dz$ measured from the AGN samples for cross-correlations in the wide redshift bin ($z\in 0.7-1.0$) for the $L_{6\mu m} > 3\times10^{44} {\rm erg\, s^{-1}}$ threshold, for the three fields of interest.}
    \label{fig:crossdNdz}
\end{figure*}

\begin{figure*}
    \centering
    \includegraphics[width = 0.99\linewidth]{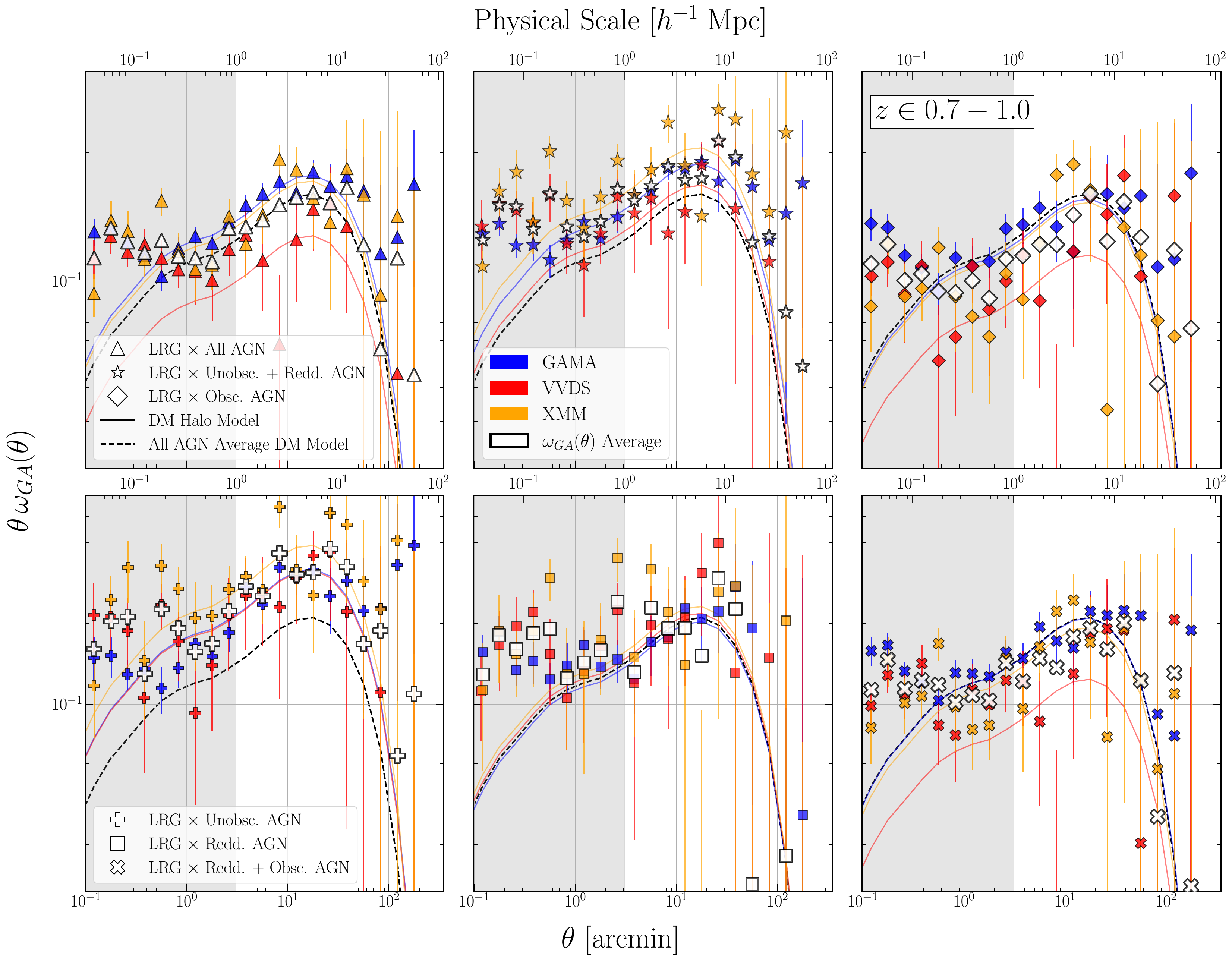}
    \caption{Cross-correlation between HSC galaxies and the full AGN sample and sub-type samples in the wide redshift bin ($z\in 0.7-1.0$) and with $L_{6\mu m} > 3\times 10^{44} \, {\rm erg\, s^{-1}}$), shown in $ \theta \omega(\theta)$. The open symbols are the per-bin averaged correlation function across fields. {The $1\sigma$ uncertainties are drawn from the square root of the diagonal of the jackknife covariance matrix for each AGN sub-type sample.} The solid lines represent the fitted \texttt{halofit} DM model to each field. The points considered in this analysis are those past the grey shaded region, at $s > 1 h^{-1} {\rm Mpc}$, where we are confident the model is able to describe the data. We note the model still agrees with the measured points well within the shaded region. We observe a departure from the DM model on small scales, as expected for the one-halo term. The dashed line represents the full AGN sample average best-fit \texttt{halofit} model, and is repeated in all panels as a reference line for ease of comparison. \textit{Top}: the cross-correlation of the full, unobscured + reddened, and obscured samples of AGN across 3 fields. \textit{Bottom}: the cross-correlation of the unobscured, reddened, and reddened + obscured samples with the LRG sample. Symbol shapes used for each sample here match those in Figure \ref{fig:all_masses}.}
    \label{fig:crosscorr_t1t2}
\end{figure*}

\renewcommand{\arraystretch}{1.3}

\begin{table*}
\centering
\caption{ Angular Correlation Function Fit Results} 
\begin{tabular}{cccccccc}
\hline \hline
\multirow{2}{*}{Subset} & \multirow{2}{*}{$N_{obj}$} & \multirow{2}{*}{Weighted $N_{obj}$}  & $\langle L_{6 \mu m} \rangle$ & \multirow{2}{*}{$\langle z \rangle$} & \multirow{1}{*}{$\langle\chi^2\rangle$} & \multirow{2}{*}{$b$} & $\langle M_h \rangle$ \\

 &  &  & [$\log$ erg s$^{-1}$]  & & [10 d.o.f.]   & & [$\log h^{-1} M_\odot$] \\

\hline \hline
 {$z\in 0.6-0.8 $} & \multicolumn{7}{c}{ } \\
\hline

LRGs & 1,288,589 & 879,258.8 & --   & $0.7\pm0.1$ & 9.8 & $1.80\pm0.03$  & -- \\
All AGN & 22,988 & 5,804.3 & $^* 44.7^{+0.4}_{-0.2}$ & $0.8^{+0.1}_{-0.2}$ & 5.4 & $1.5\pm0.1$ & $12.8\pm0.1$ \\
\hline
 {$z\in 0.8-1.0 $} & \multicolumn{7}{c}{ } \\
\hline
LRGs & 851,117 & 440,970.6 & --   & $0.9\pm0.1$ & 6.6 & $2.24\pm0.03$ & -- \\
All AGN & 26,264 & 10,381.3  & $^* 44.8^{+0.4}_{-0.2}$ & $0.9^{+0.1}_{-0.2}$ & 5.8 &$1.4\pm0.1$ & $12.7\pm0.1$\\
\hline
 {$z\in 1.0-1.2 $} & \multicolumn{7}{c}{ } \\
\hline
LRGs & 324,790 & 98,498.4 & --   & $1.0\pm0.1$ & 18.7 & $2.90\pm0.04$ & -- \\

All AGN & 25,235 & 7,204.7 & $^* 44.9^{+0.4}_{-0.3}$  & $1.1\pm0.2$ & 4.9 & $1.6\pm0.1$&  $ 12.6\pm0.1$\\

\hline
 $z\in 0.7-1.0 $ & \multicolumn{7}{c}{ } \\
\hline
 LRGs & 1,509,905 & 843,166.6 & --  & $0.8\pm0.1$ & 15.5 & $1.92\pm0.03$ & -- \\
All AGN & 28,494 & 13,898.8  & $^* 44.8^{+0.4}_{-0.2}$  & $0.9^{+0.1}_{-0.2}$ & 6.7 & $1.6\pm0.1$ &  $ 12.9\pm0.1$ \\

Unobscured AGN & 8,266 & 3,942.0  & $^* 44.9^{+0.4}_{-0.2}$ & $0.9^{+0.1}_{-0.2}$ & 8.9 & $2.2\pm0.1$ & $ 13.3\pm0.1$ \\

Unobscured + Reddened AGN & 15,156 & 6,745.71  & $^* 44.8^{+0.4}_{-0.2}$ & $0.9^{+0.1}_{-0.2}$ & 5.4 & $1.9\pm0.1$ & $ 13.1\pm0.1$ \\

Reddened AGN & 6,890 & 2,803.6  & $^* 44.8^{+0.5}_{-0.2}$ & $0.9^{+0.1}_{-0.2}$ & 3.1 & $1.6\pm0.1$ & $ 12.8^{+0.1}_{-0.2}$ \\

Reddened + Obscured AGN & 19,675 & 9,893.4  & $^* 44.8^{+0.4}_{-0.2}$ & $0.9^{+0.1}_{-0.2}$ & 5.0 & $1.4\pm0.1$ & $ 12.7\pm0.1$ \\

Obscured AGN & 12,785 & 7,089.8  & $^* 44.7^{+0.3}_{-0.2}$ & $0.9\pm{0.2}$ & 6.9 & $1.3\pm0.1$ & $ 12.6\pm0.1$ \\

\hline
High $L_{6 \mu m}$ AGN & 7,760 & 2,492.2  & $^\dagger 45.1^{+0.4}_{-0.2}$ & $1.0.^{+0.1}_{-0.2}$ & 3.6 & $1.4^{+0.1}_{-0.2}$ & $ 12.5^{+0.2}_{-0.3}$ \\
Low $L_{6 \mu m}$ AGN & 20,734 & 11,406.5 & $^\ddagger 44.8^{+0.3}_{-0.2}$ & $0.9^{+0.1}_{-0.2}$ & 6.0 & $1.7\pm0.1$ & $ 13.0\pm0.1$  \\

\hline \hline
\end{tabular}

\vspace{0.05in}

\raggedright
\footnotesize 
$^*$ Primary luminous AGN selection ($L_{6\mu m} > 3\times 10^{44}$ erg s$^{-1}$)

$^\dagger$ Higher luminous AGN selection ($L_{6\mu m} > 10^{45}$ erg s$^{-1}$)

$^\ddagger$ Lower luminous AGN selection ($3\times 10^{44} < L_{6\mu m} <  10^{45}$ erg s$^{-1}$)

\label{tab:SummMass_BinW} 
\end{table*}

\begin{figure*}
    \centering
    \includegraphics[width = 0.8\linewidth]{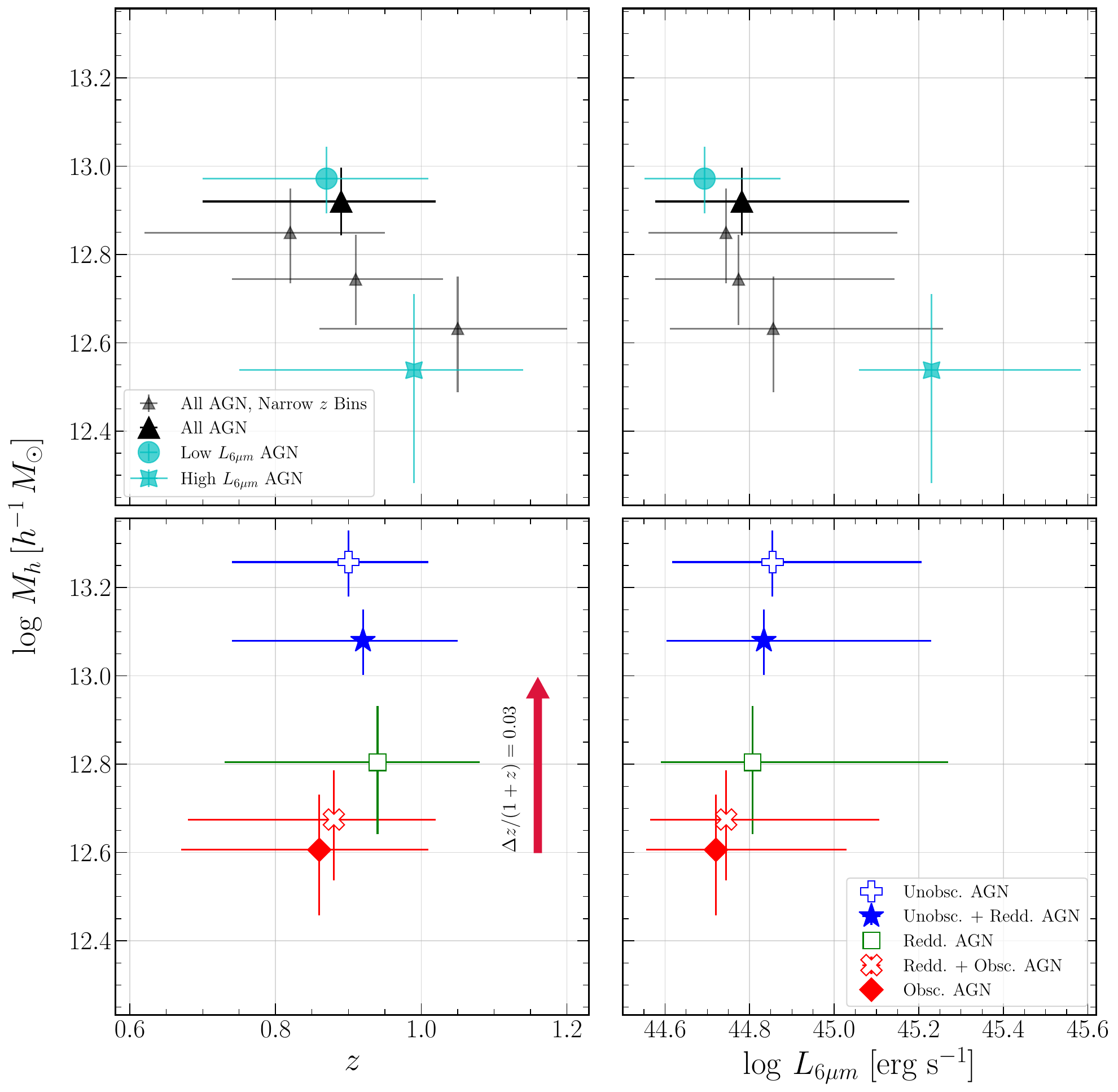}
    \caption{ Summary of recovered halo masses ($M_h$) from the AGN cross-correlation of the full AGN sample (top row), and the AGN sub-type samples (bottom row).  Left panels illustrate this as a function of redshift. Right panels show the halo masses as a function of estimated $L_{6\mu m}$. The position and uncertainties in $z$ and $L_{6\mu m}$ are drawn from the 16\%, 50\%, and 84\% quantiles of each the bin's underlying distribution. We find that the unobscured + reddened AGN are found in significantly more massive halos than obscured AGN. We additionally estimate the effect of a systematic redshift error in the $dN/dz$ on our inferred halo masses. As indicated by the red arrow, an upward $z$ shift in the $dN/dz$ by $\Delta z/(1+z) = 0.03$ would lead to a boosting of the inferred $M_h$ by magnitude of 0.4 dex, a linear factor of $\sim2.5$. {The values represented are also found in Table \ref{tab:SummMass_BinW}.}}
    \label{fig:all_masses}
\end{figure*}

In this section, we will discuss the measurement of the angular correlation functions of HSC galaxies. First we present the galaxy autocorrelations, then we turn to the AGN-galaxy cross-correlations.

\subsection{HSC Galaxy Autocorrelation}\label{sec:autocorr}

We compute the LRG autocorrelation using the methods outlined in \S \ref{sec:methods}. Our estimate of the LRG galaxy bias will be used to isolate the AGN bias from the cross-correlation in \S \ref{sec:crossall}. The $dN/dz$ for the LRGs in each redshift bin, which inform the dark matter model fit to the data, are shown in the top row of Figure \ref{fig:autodNdz}. The autocorrelations are shown in Figure \ref{fig:autocorr}. To emphasize differences between the per-field data and the models, we scale the $\omega(\theta)$ functions by a power of $\theta$. In general, recovered $\omega_{GG}(\theta)$ values agree between fields. The per-angular bin average across all fields is shown with the white symbols with a colored outline. 

As described in \S \ref{sec:model_def}, we recover the bias from these galaxy populations by fitting our \texttt{halofit} dark matter model to the measured points and uncertainties for each field. On smaller scales ($s < 0.3\,h^{-1}\,$Mpc) there is a characteristic rise in clustering that deviates in shape from our dark matter model due to multiple subhalos within the central halo (known as the ``one-halo'' term). We find the model describes the data well ($\chi^2_\nu \approx 1$) down to arcminute scales, and we limit the fits to linear scales $s > 1\,h^{-1}\,$Mpc. Fitting with $s > 2\,h^{-1}\,$Mpc yields similar results. We also test whether our results change when excluding the larger angular scales, and find that the measured bias values shift by $<1\sigma$ when fitting on scales $1<s<20 \,h^{-1}\,$Mpc. On larger scales, the error bars are significantly larger, but the measurements agree within the uncertainties.  We use the field-to-field variability in measured bias as an estimate of the total systematic uncertainty. This is exhibited in the top row of Figure \ref{fig:auto_summ}, where the fitted values of the bias from each field are plotted for a single redshift bin, and the total distribution is used to estimate the median value and its uncertainty. The recovered bias and $\chi^2_\nu$ values are summarized in Table \ref{tab:SummMass_BinW}. For the three narrow redshift bins, we find the recovered biases are (with increasing $z$) $1.80\pm0.03$, $2.24\pm0.03$, and $2.90\pm0.04$. For our wide redshift bin ($z\in 0.7-1.0$), the measured bias is $1.92\pm 0.03$. We find the reduced $\chi^2$ values for the linear bias fit to be $\lesssim 1$ in the two lower redshift bins. In the $z\in 1.0-1.2$ bin, however, we observe a significant amount of excess clustering on large scales relative to the linear DM model. Based on our preliminary galaxy clustering tests, detailed in \S\ref{sec:colorcut}, we interpret this excess as the result of projection effects from lower redshift objects still being present in our higher redshift bin. {We also perform these fits using the full covariance matrix, finding the the measured bias in two lower redshift bins shifts by $< 1 \sigma$, while the highest redshift bin has a significantly lower bias ($b_G\sim2.4$), reflective of the uncertainty in the $z\in 1.0-1.2$ bin. The constrained values and uncertainties are reported in Table \ref{tab:SummMass_BinW_covmat}.} 

We observe a characteristic rise in linear bias as a function of redshift across our three narrow redshift bins, consistent with the findings of recent LRG studies like \cite{zhou_clustering_2021}. This is expected for a population of relatively constant $M_h$ (and at a fixed magnitude limit) at greater lookback time, as has been shown by prior LRG studies \citep[e.g.][]{ishikawa_halo-model_2021}. With bias values in hand for our galaxy population, we now turn to the galaxy-quasar cross-correlations.

\subsection{Full AGN Sample Cross-Correlations}\label{sec:crossall}
Recovering the clustering amplitude from the AGN population permits us to ask questions about any trends with luminosity or redshift, and infer the halo mass that the AGN occupy following \S \ref{sec:model_def}. The cross-correlations of the LRG sample and the complete AGN catalog, $\omega_{GA}(\theta)$, are shown in Figure \ref{fig:crosscorr_all}. We perform the cross-correlation analysis for AGN that lie above a luminosity threshold of $L_{6 \,\mu m} > 3\times 10^{44} \,{\rm erg \, s^{-1}}$, as established in \S \ref{sec:lum_bins} (see bottom panels of Figure \ref{fig:autodNdz} for the samples' normalized redshift distributions). These correlation functions are calculated in each HSC field, for each of the three narrow redshift bins. We see broad agreement of the recovered $\omega_{GA}(\theta)$ across fields, but note they are noisier than the correlation for $\omega_{GG}(\theta)$. While consistent within the per-bin errors, the autocorrelations vary at $\sim 10\%$ relative to the average over $3\arcmin<\theta <40\arcmin $, and here we see $\sim 20\%$ over the same range for the lower two redshift bins, and $\sim 50\%$ variability for the highest redshift bin.

From the measured cross-correlation functions, we fit for the linear bias. Using the per-field autocorrelation bias found in \S \ref{sec:autocorr}, we divide out the galaxy bias contribution in the measured $b_G b_A$ to isolate the quasar bias, $b_A$. We perform a \texttt{halofit} model fit for scales $s > 1 {\rm h^{-1} \,Mpc}$ using CCL, and infer a halo mass as described in \S \ref{sec:model_def} above. The average bias and halo mass are recorded in Table \ref{tab:SummMass_BinW}, where they have been calculated with an inverse-variance weighting of the results from individual fields. These masses are illustrated in Figure \ref{fig:all_masses}. The recovered biases across redshift bins are consistent within $1\sigma$ (see Table \ref{tab:SummMass_BinW}), showing no significant evidence for bias evolution with redshift across this narrow redshift range. The average inferred $\log M_h$ for our $L_{6\mu m} > 3\times 10^{44} \,{\rm erg\,s^{-1}}$ range in the three narrow redshift bins is $12.8\pm0.1,\, 12.7\pm0.1$ and $12.6 \pm 0.1 $ $\log(h^{-1}\,M_\odot)$. We also calculate the correlation function for the full AGN sample in our wider redshift bin ($z\in 0.7-1.0$, Figure \ref{fig:crossdNdz}), see the upper leftmost panel of Figure \ref{fig:crosscorr_t1t2}. The inferred halo mass is $12.9\pm0.1$ $\log(h^{-1}\,M_\odot)$.

We additionally test for any evidence of halo mass evolution with luminosity by making successive subsamples with a higher threshold $L_{6 \mu m}$. The inferred mass for the $L_{6\mu m} > 10^{45}\, {\rm erg\,s^{-1}}$ sample is $12.5^{+0.2}_{-0.3}$ $\log(h^{-1}\,M_\odot)$, while the halo mass for the $ 3\times 10^{44}< L_{6\mu m} < 10^{45}\, {\rm erg\,s^{-1}}$ bin is $13.0\pm0.1$ $\log(h^{-1}\,M_\odot)$. As visualized in the upper panels of Figure \ref{fig:all_masses}, we find no significant evidence for an evolution in halo mass as a function of $L_{6\mu m}$ (also see Table \ref{tab:SummMass_BinW}). {These bias and halo mass values are consistent with the results using the full covariance matrix treatment, reported in Table \ref{tab:SummMass_BinW_covmat}.}

\subsection{AGN Sub-type cross-correlations}\label{sec:cross12}

Following the same procedure as for the full AGN sample, we calculate the cross-correlation for the AGN sub-types and estimate the physical properties of the halos in which these AGN reside. We investigate the obscured and unobscured AGN categories in a single, broader redshift bin ($z\in 0.7-1.0$) to provide a large enough sample, see total and weighted number of objects in Table \ref{tab:SummMass_BinW}. Utilizing the same $L_{6 \mu m}$ threshold, the luminosity distributions of the full unobscured + reddened and obscured AGN resemble one another, as seen in Figure \ref{fig:L6dists_W}. From a total number density of 50.1 deg$^{-2}$ ($L_{6 \mu m} > 3\times 10^{44} \,{\rm erg\,s^{-1}}$) AGN in the $z\in 0.7-1.0$ bin, there are 26.7 deg$^{-2}$ unobscured + reddened objects, and 22.5 deg$^{-2}$ obscured objects. The unobscured + reddened AGN include 14.6 deg$^{-2}$ unobscured and 12.1 deg$^{-2}$ reddened AGN. 

The measured cross-correlation functions for the complete, unobscured + reddened and obscured samples are illustrated in the top row of Figure \ref{fig:crosscorr_t1t2}, and display a fair amount of field-to-field variability, while being consistent within the errors. We account for these variations and illustrate a mean value as we do for the full AGN sample. From these correlation functions, we note how the average recovered clustering shifts above and below the reference of the complete AGN sample (black dashed line). These shifts in bias are (non-linearly) correlated with the underlying halo mass distribution of where these AGN reside. We exploit this via the methods outlined in \S \ref{sec:model_def} to recover the average mass of the halos in which these AGN are found. We again observe the rise at small angular scales ($s<0.3\, h^{-1} {\rm Mpc}$) relative to the \texttt{halofit} model from the one-halo term. As before, we fit the correlation function with the model beyond $1\, h^{-1} {\rm Mpc}$, avoiding the non-linear dominated regime. The measured correlation functions are well described by the model in our fitting range (see $\chi^2$ values in Table \ref{tab:SummMass_BinW}). {An additional test of the (in)consistency between the unobscured and obscured measured $\omega(\theta)$ values, we fit the same CCL-derived $\omega(\theta)$ DM model to both. They are fit with the (per-field) full AGN sample DM halo model (see solid lines in the top left panel of Figure \ref{fig:crosscorr_t1t2}). We find that the fitted $b^2$ value for the LRG $\times$ unobscured AGN $\omega(\theta)$ ($5.0 \pm 0.5$) is statistically inconsistent at $>5\sigma$ with the value inferred for the LRG $\times$ obscured AGN ($2.2 \pm 0.4$) measurement. This is the case when fitting with the 1-D uncertainties, as well as with the full covariance matrix. This shows that the measured amplitude of the clustering signal from both of these samples is substantially different.} 

With these correlation functions, we seek to measure the clustering strength of our different AGN sub-samples. In turn, we estimate the galaxy bias and inferred halo mass for each sub-type to investigate whether the different samples have the same or different characteristic $M_h$. We present the AGN sub-type halo masses in Fig \ref{fig:all_masses}, in Table \ref{tab:SummMass_BinW}{, and we present the full covariance matrix treatment's inferred $M_h$ and significance in parentheses throughout the text and in \S \ref{sec:app_covmat} in Table \ref{tab:SummMass_BinW_covmat}. The $M_h$ values inferred with the full covariance matrix treatment shift by $< 1\sigma$ for all AGN sub-types.} The average unobscured + reddened AGN halo mass is $13.1\pm 0.1 \, (13.2^{+0.1}_{-0.2})$ $\log(\, h^{-1}\,M_\odot)$. We find the $M_h$ for obscured AGN is $12.6\pm0.1 \,(12.6^{+0.2}_{-0.3})$ $\log(\, h^{-1}\,M_\odot)$. The unobscured + reddened average $M_h$ is higher than the obscured halo mass by a factor of $\sim 3$. Given the 1-D statistical uncertainties, this is a $3 \sigma$ difference {($2.8\sigma$ with the full covariance matrix)}.

We also investigate what happens to the clustering and the inferred halo mass when we split the unobscured + reddened AGN sample into its reddened and unobscured AGN components. The correlation functions for these samples can be seen in the bottom panel of Figure \ref{fig:crosscorr_t1t2}.  First we compare the average halo mass for the unobscured and reddened AGN, finding $\log M_h = 13.3\pm 0.1 \,(13.4\pm 0.1)$ and $ 12.8^{+0.1}_{-0.2} \,(13.0^{+0.1}_{-0.2})\,\log(\, h^{-1}\,M_\odot)$, respectively. Herein, our sample of unobscured objects are more massive than their reddened counterparts by a factor of $\sim 3$, at $3\sigma$ {($2.9 \sigma$ with the full $C_{jk}$)}. Next we contrast blue objects (unobscured AGN) and reddened objects (reddened AGN + obscured AGN), finding that there is a significant difference, with the former being $\sim 4\times$ more massive, at $5\sigma$ {($3.9 \sigma$)}. We directly compare the unobscured and obscured AGN, and find a $5\sigma$ {($3.5 \sigma$)} statistical difference in the inferred average halo mass, where the unobscured AGN reside in $\sim 4.5\times$ more massive halos.

These measurements may suggest that the AGN populations with different obscuration levels occupy halos with different average inferred masses. However, we must consider possible systematic uncertainties in our photometric redshift values and possible misclassifications before making any physical interpretations.

\subsection{Potential Effects of Unconstrained $dN/dz$ Uncertainties} \label{sec:dndzshift}

We investigate how systematic shifts in our $dN/dz$ would impact the interpretation of our measured clustering functions. As we detailed in \S \ref{sec:model_def}, the halo model is directly informed by the overlap of the input $dN/dz$'s in a cross-correlation. The smaller the overlap of the two samples' $dN/dz$, the lower the expected clustering amplitude of the halo model will be. Consequently, the measured bias relative to the halo model will be higher if these $dN/dz$ are shifted away from each other. 

Here we use a toy scenario to estimate the magnitude of the required shift in $dN/dz$ to resolve the currently observed difference in average halo mass estimated between our unobscured + reddened and obscured AGN. We shift the $dN/dz$ for our AGN samples and subsamples for each field by $\Delta z/ (1+z) = \pm 0.03$. We find that shifting the $dN/dz$ to lower redshifts has a minimal impact on the measured bias or inferred $M_h$. This is expected given the shift to lower redshift does not significantly change the overlap in $dN/dz$ between the LRG and the AGN. Shifting the distribution to higher $z$ (further separating it from the bulk of the LRG $dN/dz$), shifts the average inferred halo mass up by a factor of $\sim 2.5$ ($0.4$ dex) for any sample/subsample of AGN. This shift is represented by the red arrow in the bottom left panel of Figure \ref{fig:all_masses}. The $b_A$ is shifted higher by $0.6, 0.3$ for the unobscured + reddened, and obscured samples, respectively. 

If the true $dN/dz$ for the obscured objects were to be systematically shifted in this way relative to our measurement (while the unobscured + reddened objects' $dN/dz$ is accurate), then the significant difference we measure here would be resolved. As \cite{hickox_clustering_2011} show, systematic shifts in the redshifts for photometrically-determined obscured AGN has been a topic of considerable uncertainty given the abundance of spectroscopic information on unobscured AGN \citep[c.f.][]{shen_quasar_2009, lyke_sloan_2020} and the relative dearth of obscured AGN spectra. {This is sensible given obscured objects are fainter and will suffer similar photo-$z$ fitting degeneracies as galaxies at similar redshifts.} Prior studies from HSC have also highlighted the possibility of these systematic shifts in the photometric redshift distributions \citep[c.f.][]{dalal_hyper_2023}. 

{ We also investigate the effect of having uncorrelated objects (any object that is at a different redshift but is incorrectly in our redshift bin) in the analysis. We estimate how including increasing numbers of randomly distributed objects (i.e. with incorrect redshifts) in the unobscured and obscured AGN samples affects the measured correlation functions. We find that if $10\%$ of the unobscured AGN sample were in fact uncorrelated random objects (that are then unaccounted in the $dN/dz$), it would depress the inferred bias value such that it would resolve the difference in bias with the obscured AGN. The magnitude of this effect is reproduced when including random objects in the obscured AGN sample. We infer from this test that if we had a similar fraction of catastrophic photo-$z$ misattributions only in our obscured AGN sample, it could produce the bias difference we measure in this analysis. Clearly, photometric redshifts provide a systematics floor to our results that can only be resolved with large spectroscopic samples. We discuss the possibility of these redshift failures and other potential contaminating sources in our sample in \S\ref{sec:T1T2}}.

\section{Discussion} \label{sec:disc}

\begin{figure*}
    \centering
    \includegraphics[width = \linewidth]{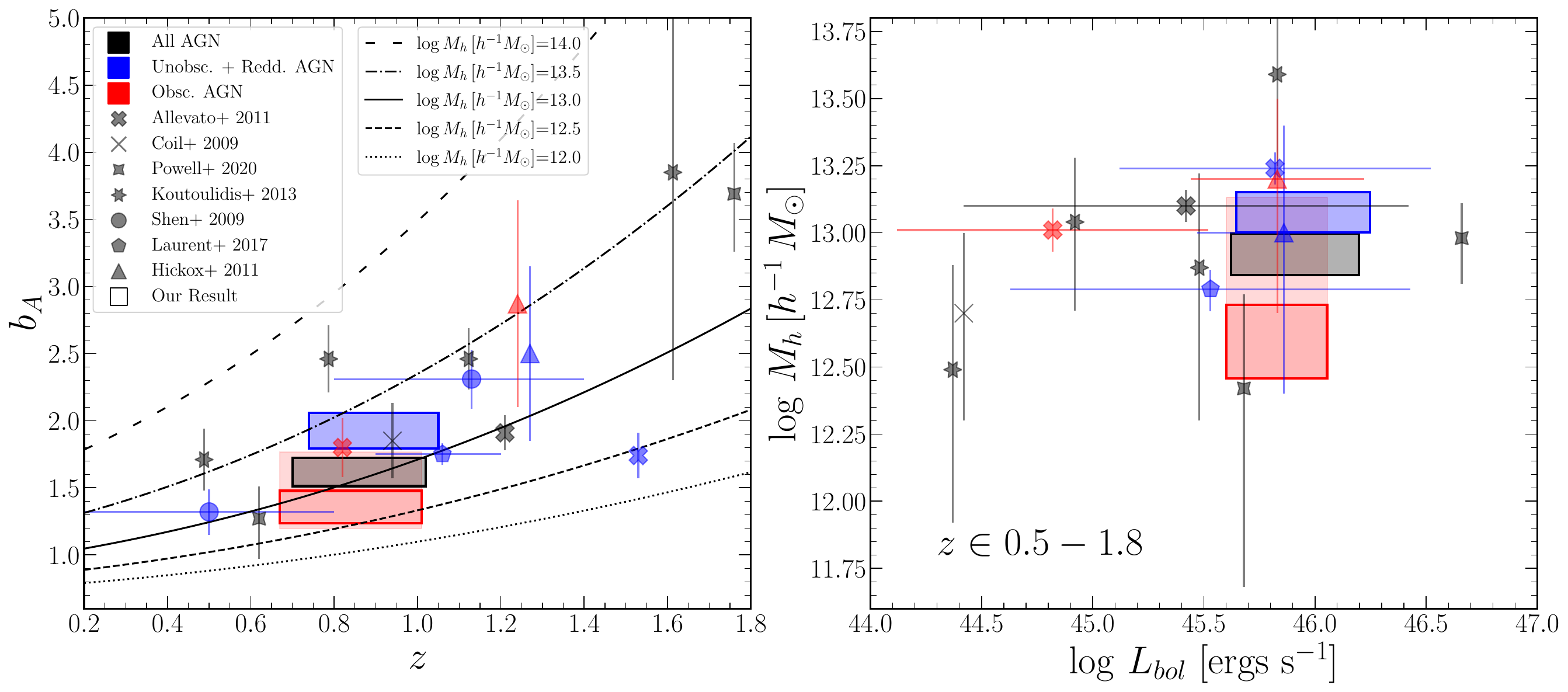}
    \caption{Our measured AGN bias and inferred halo masses for the full AGN sample and unobscured + reddened and obscured samples in comparison with other analyses. 
    \textit{Left:} The recovered AGN bias as a function of redshift in comparison with X-ray, optical, and IR selected AGN clustering studies. Marker styles represent the analysis used in comparison. Marker color indicates whether it is an analysis of all the AGN in their sample, or a subsample for unobscured + reddened or obscured AGN. The colored boxes illustrate our measurements and statistical uncertainties (solid outlines), and we include our estimate of the possible magnitude of inferred bias and halo mass shifts were there to be a systematic bias in the $dN/dz$ for $\Delta z/(1+z)= 0.03$ (lower opacity red rectangle). Plotted lines illustrate the nominal $b(z, M_h)$ tracts for different halo masses \citep{tinker_large-scale_2010}. 
    \textit{Right:} Inferred halo masses as a function of bolometric luminosity ($L_{bol}$), markers are the same as in the left panel. We convert studies' recorded $L_\nu$ to a $L_{bol}$ using the correction from \cite{hopkins_observational_2007}.
    We compare, where available, with values from projected angular correlation function analyses, rather than projected real-space derived values \citep[cf.][]{hickox_clustering_2011}. This halo mass comparison serves as a visualization of other results in the field and the halo mass differences between sub-types they find, but absolute value comparisons are difficult since studies presented here use different $b-M_h$ formulations.}
    \label{fig:comparison_plot}
\end{figure*}

The use of correlation functions has become an efficient means of investigating the properties of dark matter halos in which AGN reside. From illustrating redshift evolution of the bias for AGN samples \citep[cf.][]{shen_quasar_2009, ross_clustering_2009, allevato_xmm-newton_2011, koutoulidis_clustering_2013, powell_clustering_2020}, to the bias and inferred halo mass of different AGN sub-types \citep[e.g.,][]{hickox_host_2009, allevato_clustering_2014, dipompeo_updated_2016, dipompeo_characteristic_2017, laurent_clustering_2017, petter_host_2023, viitanen_large-scale_2023}, spatial statistics have served to describe underlying properties of the distribution and abundance of AGN.

We study the halo properties of our HSC + WISE AGN sample using cross-correlation with LRGs. We find that the AGN population shows no evidence for halo mass evolution with redshift, over the narrow redshift range studied here. Focusing on AGN with $L_{6 \mu m} > 3\times 10^{44}$ erg s$^{-1}$ where we are relatively complete up to $z\sim 1.2$, we likewise find no evidence for variation in $M_h$  as a function of luminosity. This is equivalent to $L_{bol} >3.4\times10^{45}$ erg s$^{-1}$, following the IR correction from \cite{hopkins_observational_2007}. However, we do see a dramatic difference in the inferred halo mass between the unobscured + reddened and obscured AGN samples. The masses differ by a factor of $\sim3$, with the unobscured + reddened AGN being more massive at $>3\sigma$ significance relative to the obscured objects. 

\subsection{Luminosity Dependence of Inferred Halo Properties}\label{sec:disscussion_1}

Many prior papers have found a similar lack of luminosity dependence in clustering as found here. Direct comparisons between samples can be complicated by differences in selection method, redshift distributions, and so on. We thus limit our attention to samples with $\langle z\rangle \approx 1$, with photometric or spectroscopic redshifts, and that utilize angular correlation functions. We compare with X-ray \citep{coil_aegis_2009, allevato_xmm-newton_2011, koutoulidis_clustering_2013, powell_clustering_2020}, optical/UV \citep{shen_quasar_2009, laurent_clustering_2017}, and MIR \citep{hickox_clustering_2011} selected samples. The left panel of Figure \ref{fig:comparison_plot} includes bias tracks as a function of redshift for a given halo mass, following \cite{tinker_large-scale_2010}. The right panel shows the reported $M_h$ from these analyses as a function of bolometric luminosity. To convert each sample to a bolometric luminosity in a uniform manner, we use the corrections from \cite{hopkins_observational_2007}. We find the average halo mass for our AGN sample ($\approx 13\,\log h^{-1} {\rm M_\odot}$) is consistent with the AGN halo mass inferred by the studies we compare with here.

We reiterate that comparing halo mass estimates across analyses is imperfect due to different treatments of the galaxy bias to halo mass connection, which may introduce significant shifts. Attention must be paid to whether the inferred $M_h$ is calculated with either the effective redshift or the complete $dN/dz$ of the sample, as well as the precise $b-M_h$ connection used \citep[e.g.][]{sheth_large_1999, sheth_ellipsoidal_2001, tinker_large-scale_2010}. Moreover, distinct cosmological codes make different assumptions in standard galaxy bias$-$halo mass treatments, such as the mass definition adopted, which in the case of spherical overdensity-based masses picks a value of $\Delta$ as defined in \cite{tinker_large-scale_2010}. We have assumed, like \cite{laurent_clustering_2017} and in the CCL default \citep{chisari_core_2019},  that $\Delta = 200$. Additionally, there is not an agreed upon convention for the definition of halo mass (and therefore, a universal value of $\Delta$). This plurality of different approaches can introduce significant shifts on the order of $0.1-1.0$ dex, and makes any comparison across analyses difficult. To ensure robustness in our investigation, we check our halo mass inference from (the $N$-body derived) \cite{tinker_large-scale_2010} against the analytic formulation from \cite{sheth_large_1999}. While the recovered halo masses for all our sub-samples shift to $\sim0.2$ dex larger values, we find that the relative halo mass differences between AGN sub-types are consistent across these $b-M_h$ parametrizations. {We also verify that these halo mass differences are preserved when choosing a different halo mass function formalism. We chose a parameterization that integrates over the halo mass function so as to recover the $\langle M_h \rangle$ ( see \cite{laurent_clustering_2017} for another implementation of this method). Other analyses such as \cite{petter_host_2023} have opted for an $M_{\rm eff}$ approach where one reports the halo mass at which the $b(z,M)$ matches the measured galaxy bias, for a given $b-M_h$ connection. While there are physical implications to each of these approaches, we find that the estimated $M_h$ differences and significances are present irrespective of the chosen formalism. We are not so focused on the precise value of the inferred $M_h$, but rather the significance of the differences. Given all these possible analysis choices, w}e conclude that while relative halo masses within any investigation are informative, contrasting absolute values between analyses can be misleading.

As we noted previously, we find no evidence for bias or halo mass evolution as a function of luminosity. Our result is consistent with \cite{shen_quasar_2009}, who find that over a wide range of redshifts, the recovered halo mass of optically-selected quasars does not show any trends with luminosity \citep[see also,][]{croom_2df_2005, lidz_luminosity_2006, coil_aegis_2009, hickox_clustering_2011, mendez_primus_2016}. Similarly, \cite{shen_cross-correlation_2013} find there is a poor correlation between $L$ and $M_h$ at $\bar{z} \sim 0.5$.

Several analyses have constrained the AGN halo mass range within $M_h \in 12.5 - 13\,\log h^{-1}\,M_\odot$ \citep{hopkins_cosmological_2008, cappelluti_clustering_2012, krumpe_more_2014, timlin_clustering_2018}. It has been suggested that the halo mass scale at which the studies presented here converge is unique for AGN triggering due to the low relative galaxy velocities within a group \citep[e.g.,][]{hopkins_cosmological_2008, hickox_clustering_2011}. However, studies of more detailed semi-empirical models have suggested that the underlying distribution of halo masses is broad, but the combination of higher AGN fractions in higher mass star-forming galaxies, low-$L$ AGN sample incompleteness, and relatively high satellite fraction among AGN leads to an apparent constant $M_h$ \citep{georgakakis_exploring_2019, aird_agn-galaxy-halo_2021}.

We note that some analyses have found evidence for a correlation between quasar luminosity and inferred halo mass \citep{koutoulidis_clustering_2013}. \cite{krumpe_spatial_2018} study X-ray AGN at $\langle z \rangle < 0.3$. They find no significant difference between their high and low $L_X$ bins on linear scales (two-halo contribution), but find a significant difference (3$\sigma$) when comparing the HOD-derived clustering strength for the one-halo term. This difference may be driven by the inclusion of the intra-halo small scale clustering (the one-halo term), given that prior studies have noted this higher clustering amplitude due to the presence of merging systems \citep{hennawi_binary_2006, serber_small-scale_2006, hopkins_cosmological_2008}. Since our analysis only considers large scales, we are not surprised at the lack of a luminosity difference in our results. This absence of a $L$ dependence allows us to analyze differences as a function of AGN type (\S \ref{sec:T1T2}), without requiring identical $L$ distributions.

\subsection{Inferred Average Halo Mass Differences} \label{sec:T1T2}

Previous studies have disagreed on whether obscured/Type II and unobscured/Type I AGN reside in dark matter halos of different masses. For instance, \cite{allevato_xmm-newton_2011} find a trend where (X-ray selected, optically classified) broad-line AGN are in more massive halos than narrow-line AGN at a range of redshifts ($z_{\rm spec} \sim0.8-2.0$), and repeat this finding at $z_{\rm spec} \sim 3$ in \cite{allevato_clustering_2014}. Nevertheless, it is difficult to compare with X-ray analyses given smaller sample sizes and different redshift ranges. \cite{krumpe_spatial_2018} find excess clustering of (X-ray column depth-defined) Type II AGN relative to Type I AGN using spectroscopic redshifts for $ \langle z_{\rm spec}  \rangle <0.04$. \cite{cappelluti_active_2010} found the opposite, showing X-ray Type I AGN to be an order of magnitude more massive than their Type II counterparts for $\langle z_{\rm spec} \rangle <0.06$, again split by their column depth. 

Previous optical and IR-selected AGN studies also see a divergence of results. Using a projected real-space correlation analysis of optical$-$MIR color selected quasars, \cite{hickox_clustering_2011} show that the obscured AGN (with primarily photometric redshifts) showed statistically consistent clustering with the unobscured AGN (with primarily spectroscopic redshifts), but noted that the difference may be underestimated due to photo-$z$ uncertainties. \cite{mendez_primus_2016} also find no significant difference between AGN sub-types across IR selections with spectroscopic redshifts for each object. Other IR analyses have focused on using representative $dN/dz$ distributions from a subset of their IR-selected AGN sample, which complicates direct comparison. \cite{petter_host_2023} find, as do previous photometric AGN surveys from IR selection \citep{donoso_angular_2014, dipompeo_characteristic_2017}, that obscured AGN are in significantly higher mass halos than their unobscured counterparts. But comparisons with these studies are not straightforward since many of these analyses \citep[such as ][]{gilli_spatial_2009, donoso_angular_2014, dipompeo_updated_2016, dipompeo_characteristic_2017, petter_host_2023} span redshift ranges ($z\in 0-3$) that we do not probe in this paper. In addition to differences in redshift range and approach, we also utilize a different color selection, which also precludes direct comparisons.

We find that the $g-$W3 color vs. redshift selected unobscured + reddened AGN in our sample are found in more massive halos than obscured AGN, by a factor of $\sim 3$, with a statistical significance $>3\sigma$ {($2.8 \sigma$ with the full covariance matrix)}. We also find that the host halos of unobscured AGN are more massive than the halos of reddened AGN by a factor of $\sim 3$, at a statistical significance of $3\sigma$ {($2.9 \sigma$ with the $C_{jk}$)}, and the unobscured AGN are $\sim 4.5\times$ more massive than the obscured AGN at $5\sigma$ {($3.5\sigma$), where the central values of the inferred AGN sub-type $M_h$ shift by $<1\sigma$ in all cases.} We note that the systematic uncertainty in our inferred average $M_h$ associated with our reliance on photometric redshifts can be as large as the significant difference we see between our sub-samples with a $\Delta z\sim+0.06$ systematic shift in the $dN/dz$  (as visualized by the low opacity red rectangle in Figure \ref{fig:comparison_plot}). We cannot rule out that such small systematic shifts in the $dN/dz$ could drive our and previous (photo-$z$ based) results regarding $M_h$ for obscured vs unobscured AGN (see \S \ref{sec:dndzshift}). 

We leave open the possibility that some objects can be misclassified and included in our AGN sample. Objects such as lower-mass star-forming galaxies \citep{hainline_mid-infrared_2016} could enter our selection, but only if the redshift solution is incorrect such that the estimated luminosity is above our threshold, and if it was sufficiently bright in the MIR to appear to be similar to an AGN. These tight observational constraints significantly limit the parameter space from which non-AGN may enter our sample. Though we are confident our classification methods are robust, further observations are key to reducing the possibility of this potential systematic effect.

We conclude that additional spectroscopic measurements are necessary to determine whether the differences in halo mass by obscuration are real. If so, we may have found evidence linking obscuration level with different phases in AGN-host galaxy coevolution, as others have proposed. Such differences may arise naturally if (for instance) earlier phases of galaxy and black hole growth are associated with lower-mass halos and higher levels of obscuration \citep[e.g.,][]{hopkins_observational_2007, hickox_host_2009, fawcett_striking_2023}. While photometric surveys provide large sample statistics from which to perform clustering measurements, spectroscopic information is key for robust AGN identification, classification, and redshift determination. Large format spectroscopic surveys, like the upcoming Prime Focus Spectrograph \citep{takada_extragalactic_2014, greene_prime_2022} on the Subaru Telescope, will be essential in further investigating the ensemble properties of AGN and reduce the effect of systematics.

\section{Conclusions}\label{sec:conclu}

We present a correlation analysis between luminous red galaxies observed by HSC and active galactic nuclei selected from HSC and WISE photometry at angular scales of $0.1\arcmin < \theta < 200\arcmin$. These AGN are selected with a combination of HSC optical and WISE MIR photometric colors, and their classification has been shown to be robust with spectroscopic confirmation \citep{hviding_spectroscopic_2024}. Using three HSC fields totaling $\sim 600 \,{\rm deg^{2}}$, we have a total of $1.7 \times 10^6$ LRGs and $\sim 34,000$ AGN in the full redshift and luminosity range we analyze($z\in 0.6-1.2$, $L_{6\mu m} > 3\times 10^{44} {\rm \, erg\,s^{-1}}$. For the AGN sub-type cross correlation clustering analysis, we use a single redshift bin ($z\in 0.7-1.0$) containing $1.5\times 10^6$ LRGs and $\sim 28,500$ $L_{6\mu m}$-limited AGN. We fit these correlation functions with a linear+non-linear DM halo model at physical scales $s > 1\, h^{-1}{\rm Mpc}$ ($\theta \gtrsim 3\arcmin$), and interpret the clustering strength with physical parameters. Our principal conclusions are as follows.

\begin{enumerate}
    
    \item We find no significant evidence for luminosity dependence on the inferred halo mass where AGN reside.
    \item The host halos of unobscured + reddened AGN are $\sim3\times$ more massive than those of obscured AGN, at a $3\sigma$ statistical difference {($2.8\sigma$ with the full covariance matrix)}. We also directly compare our sample of unobscured AGN (i.e. Type I quasars) and obscured AGN (i.e. Type II) and find the former are, on average, found in $\sim4.5\times$ more massive halos with $\log M_h = 13.3\pm0.1\, \log\, h^{-1} M_\odot$, a $5\sigma$ statistical difference. {If we use the full covariance matrix instead, we find that the $M_h$ values are unchanged, with a $3.5\sigma$ statistical difference between unobscured and obscured AGN.} 
    
    \item We find that reddened AGN (that we expect to have broad lines) are found in halos of intermediate mass between unobscured and obscured AGN, at $\log M_h = 12.8^{+0.1}_{-0.2}\, \log\, h^{-1} M_\odot$. As such, the halos of unobscured objects are $\sim3 \times$ more massive than reddened AGN, at $3\sigma$ (given statistical uncertainties {from the diagonal of the covariance matrix}). This result requires additional spectroscopic follow-up to better characterize the AGN samples, but could point to an evolutionary sequence between these AGN that is being traced by the average halo mass. 
    
\end{enumerate}

We investigate the coevolution of SMBHs and their galactic hosts, and find that inferred average halo masses continue to be an effective means of tracing average AGN properties. By holding the luminosity and redshift distributions relatively constant between AGN subsamples, we are able to infer the differences between AGN sub-types as a function of our photometric color and redshift classification. However, possible unconstrained $dN/dz$ systematic uncertainties prevent us from concluding that these inferred differences represent the underlying distribution with certainty, if they were to only affect the lower halo mass results. HOD analyses are essential to exploit the small scale clustering measurements we have shown here, and are a promising extension of this study to learn about AGN satellite fractions. Additional work is necessary to continue constraining the properties of the obscured AGN, but the results here show that it may include a significant population of lower halo-mass objects, relative to the unobscured + reddened AGN. Evolutionary models and semi-empirical approaches like those outlined by \cite{aird_agn-galaxy-halo_2021} would be useful in constructing simulations where we may test for possible halo mass differences. Inferring their characteristics has been an area of active research \citep{petter_host_2022, fawcett_striking_2023}, and further observational data is key to understand these AGN's morphology and potential place in an AGN evolutionary picture. 

The data underlying this analysis are set to improve significantly in the coming years. Upcoming wide-field photometric surveys like the Rubin Observatory's Legacy Survey of Space and Time \citep{ivezic_lsst_2019} and eRosita \citep{merloni_erosita_2012}, and spectroscopic surveys such as the Prime Focus Spectrograph \citep{takada_extragalactic_2014, greene_prime_2022} and the Dark Energy Spectroscopic Instrument \citep{desi_collaboration_desi_2016, desi_collaboration_desi_2024} will be particularly capable of revealing AGN (and their sub-types') populations. From disentangling AGN properties as a function of time and evolutionary stage, to better characterizing the spectral properties of AGN and the galaxies in which they reside, new datasets will readily enable extensions of the work presented here.

\section*{Acknowledgments}

We thank {the anonymous referee for their many helpful comments. We thank} R. Lupton, {P. Melchior}, N. E. Chisari, and N. Bahcall for helpful conversations throughout the course of this work. 

Computing was performed using the Princeton Research Computing resources at Princeton University. RCR acknowledges support from the Ford Foundation Predoctoral Fellowship from the National Academy of Sciences, Engineering, and Medicine. ADG and RCR gratefully acknowledge support from the NASA Astrophysics Data Analysis Program \#80NSSC23K0485. ADG and JEG acknowledge support from the National Science Foundation under Grant Number AST-1613744{, and JEG acknowledges support from the National Science Foundation under Grant Number AST-2306950.}

The Hyper Suprime-Cam (HSC) Collaboration includes the astronomical communities of Japan and Taiwan, and Princeton University. The HSC instrumentation and software were developed by the National Astronomical Observatory of Japan (NAOJ), the Kavli Institute for the Physics and Mathematics of the Universe (Kavli IPMU), the University of Tokyo, the High Energy Accelerator Research Organization (KEK), the Academia Sinica Institute for Astronomy and Astrophysics in Taiwan (ASIAA), and Princeton University. Funding was contributed by the FIRST program from the Japanese Cabinet Office, the Ministry of Education, Culture, Sports, Science and Technology (MEXT), the Japan Society for the Promotion of Science (JSPS), Japan Science and Technology Agency (JST), the Toray Science Foundation, NAOJ, Kavli IPMU, KEK, ASIAA, and Princeton University. 

This paper makes use of software developed for Vera C. Rubin Observatory. We thank the Rubin Observatory for making their code available as free software at \url{http://pipelines.lsst.io/}. This paper is based on data collected at the Subaru Telescope and retrieved from the HSC data archive system, which is operated by the Subaru Telescope and Astronomy Data Center (ADC) at NAOJ. Data analysis was in part carried out with the cooperation of Center for Computational Astrophysics (CfCA), NAOJ. 

This publication makes use of data products from the Wide-field Infrared Survey Explorer, which is a joint project of the University of California, Los Angeles, and the Jet Propulsion Laboratory/California Institute of Technology, funded by the National Aeronautics and Space Administration.

This publication also makes use of data products from
NEOWISE, which is a project of the Jet Propulsion
Laboratory/California Institute of Technology, funded by the
Planetary Science Division of the National Aeronautics and
Space Administration.

We are honored and grateful for the opportunity of observing the Universe from Maunakea, which has the cultural, historical and natural significance in Hawaii. 

\facilities{Subaru (HSC), WISE, NEOWISE, Sloan}

\software{\texttt{Astropy} \citep{astropy_collaboration_astropy_2018, astropy_collaboration_astropy_2022}, \texttt{Matplotlib} \citep{hunter_matplotlib_2007}, \texttt{NumPy} \citep{van_der_walt_numpy_2011, harris_array_2020}, \texttt{SciPy} \citep{virtanen_scipy_2020}}, \texttt{Corrfunc} \citep{sinha_corrfunc_2020}, Core Cosmology Library \citep{chisari_core_2019}

\bibliographystyle{aasjournal}
\bibliography{references}

\appendix

\section{LRG Selection}\label{sec:colorcut}

\begin{figure*}
    \centering
    \includegraphics[width=\linewidth]{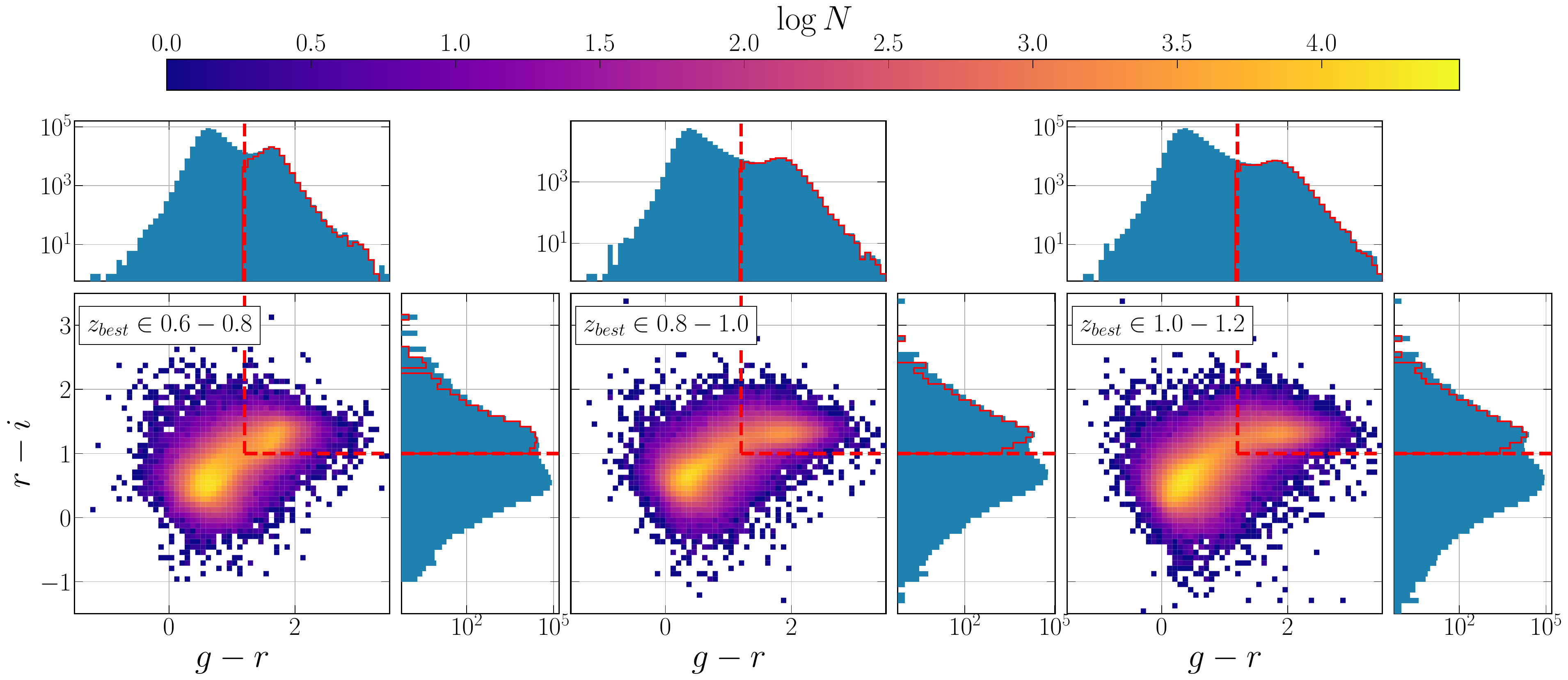}
    \caption{HSC Color-Color Diagram for the HSC galaxy sample in the XMM field, across three narrow redshift bins. As described in \S \ref{sec:colorcut}, we isolate an LRG sample from the whole galaxy catalog via a color-color cut, $g-r > 1.2$ and $r-i > 1.0$ (as indicated by the red dashed lines). The objects are isolated in a specific redshift bin given the fiducial redshift ($z_{best}$) for every object. The histograms illustrate the distribution of galaxies in color space, and illustrate the scale at which we separate the LRG population from rest of the galaxies. The red solid lines in the histograms highlight the color distributions of the final selection of objects.}
    \label{fig:XMM_colorcolor}
\end{figure*}

We have defined our parent HSC galaxy samples as objects brighter than $i=24$ AB mag, using three narrow photometric redshift bins, and a wide redshift bin. We determined the bins widths based on the median uncertainty from the AGN sample's photometric redshifts (see \S \ref{sec:zbins}). As a check on the photometric redshift accuracy, we calculate the cross-correlation between the galaxy sample in first and third narrow redshift bins (i.e. before color cuts). If our photometric redshifts and their errors are reliable, we should not detect much signal above random in the cross-correlation. Nonetheless, by estimating the amplitude of the cross-bin correlation relative to the autocorrelations, we find that there is $\sim 30\%$ correlation between the full magnitude-limited galaxy samples in the 0.6-0.8 and 1.0-1.2 redshift bins. In addition, the autocorrelation of this sample displays artificial signal on large scales, where the true signal is about $1\%$ of the small scale amplitude and we become systematics limited. These findings require that we prune the galaxy catalog and identify the source of these systematic limitations to our measurement. 

To fix this, we adopt a color-space sample selection to isolate luminous red galaxies (LRGs).
LRGs have been found to have more reliable redshifts, including prior HSC analyses \citep{eisenstein_spectroscopic_2001, rau_weak_2023}. This is primarily due to the presence of the $4000\,\AA$ Balmer break in the galaxy SED, characteristic of LRGs, which allows photometric redshift codes to find a more stable solution. Using the measured \texttt{CModel} AB magnitude for HSC galaxies, we select objects with  $g-r > 1.2$ and $r-i > 1.0$ colors to isolate the LRG's readily seen in color-color diagrams, shown in the color-space representation of the galaxies in the XMM field in Figure \ref{fig:XMM_colorcolor}. Given the HSC photometric bands, this color selection isolates LRGs whose redshifts are close to the $0.7\lesssim z \lesssim 1.0$ range.  We find the number of objects is reduced significantly, removing between $80\%$ and $95\%$ of the total galaxy sample (for the lowest and highest redshift bins, respectively). The inferred stellar mass available in the \texttt{Mizuki} catalog for the complete ($i > 24$) galaxy population peaks at $\sim 9 \times 10^{9} M_\odot$, while for the red galaxy sample the peak is at $\sim 6 \times 10^{10} M_\odot$. After this cut, the cross-correlation between the red galaxy samples in the lowest ($z\in 0.6-0.8$) and highest ($z\in 1.0-1.2$) redshift bins is $\sim 10\%$ the amplitude of the autocorrelation in a single bin. We find this reduction in systematic uncertainty to be sufficient for our analysis and propagate the color cut throughout the rest of the experiment. 

The per redshift bin LRG galaxy sample $dN/dz$ is illustrated in the top row of Figure \ref{fig:autodNdz}. The multiple peaks in each redshift bin are reflective of unphysical fitting degeneracies in the HSC photometric redshift code, wherein certain values are over-prescribed. There are also secondary peaks outside the redshift bin after the weighted addition, indicative of a significant number of objects whose $p(z)$ is clustered outside (but significantly overlaps with) the bin. We test how these $dN/dz$ shapes affect our forward model, and find a Gaussian-smoothed version of these distributions produce forward models that diverge from the fiducial model by $< 0.5\%$. These distributions are representative of the estimated redshifts and their uncertainties in each bin.

\section{Proper Accounting of $p(z)$ Uncertainties} \label{sec:weightedmethod}

The standard tomographic method for redshift binning takes into account only those galaxies whose fiducial redshift is in the bin. This procedure would, in a given tomographic redshift bin, exclude objects whose nominal redshift ($z_{best}$) was not inside the bounds of the bin, even if a significant fraction of their $p(z)$ did fall in the bin. We seek to account for the complete redshift uncertainty in our sample, for objects that both scatter into and out of our redshift bins. As such, we implement the following formalism to create a more representative clustering statistic taking this into account. 

\subsection{Redshift-Weighted Correlation Function}
 We define a weight $\mathcal{W}_i$ based on the fraction of an object's $p(z)$ found within a bin
\begin{equation}
    \mathcal{W}_i = \frac{\int^{b}_{a} p_i(z) \, dz }{\int_0^{\infty} p_i(z)\, dz}
\end{equation}
for a redshift bin with edges $a$ and $b$. We define our binning selection such that $\mathcal{W}_i \geq \mathcal{W}_0$, where $\mathcal{W}_0$ is the threshold we establish for inclusion in our redshift bin. In this analysis, any object with greater than $3\%$ of the probability $p(z)$ falling inside the predefined bin is included. This choice reflects a balance of the inclusion of all available redshift space while not making the calculation intractable.

We implement this weight for a pair counting operation such that

\begin{equation}
    \mathcal{W}_{XY} (\theta) = \frac{\sum_i^{N_X} \sum_{j \neq i}^{N_Y} \mathcal{W}_{i,X} \mathcal{W}_{j,Y} }{\sum_i^{N_X} \sum_{j \neq i}^{N_Y} \bar{\Theta}_{ij,k}}
\end{equation}
and substitute the total number of objects for the summed weights
\begin{equation}
    \mathcal{W}_{X} = \sum_i^{N_X} \mathcal{W}_{i,X}.
\end{equation}
The correlation function element is thus defined as 
\begin{equation}
    XY(\theta) = \frac{\mathcal{W}_{XY}(\theta) X Y^\prime(\theta)}{\mathcal{W}_{X} \mathcal{W}_{Y}}.
\end{equation}
We substitute this expression into the \cite{landy_bias_1993} estimator in Equation \eqref{eq:LS} to account for appropriate object weighting.

\subsection{Redshift-Weighted Forward Modeling}
We also update the $dN/dz$ calculation to include the appropriate per-object weighting where 
 \begin{equation} \label{eq:wdNdz}
    \frac{dN}{dz} = \sum_i \mathcal{W}_i \cdot p_i(z~|~\mathcal{W}_i \geq \mathcal{W}_0) .
\end{equation}
These are used in the forward model as described in Section \ref{sec:model_def}. 

Following significant testing and comparison with standard tomographic binning, we find that the two methods return values that are entirely consistent ($<1\sigma$) with each other. Nevertheless we implement our weighted analysis throughout this work, preferring to fold in all the available $p(z)$ information. In turn, this method is less subject to systematic biases inherent in photometric redshift fitting.

\section{All AGN Sub-type $dN/dz$}

\begin{figure}[h!]
    \centering
    \includegraphics[width=1.\linewidth]{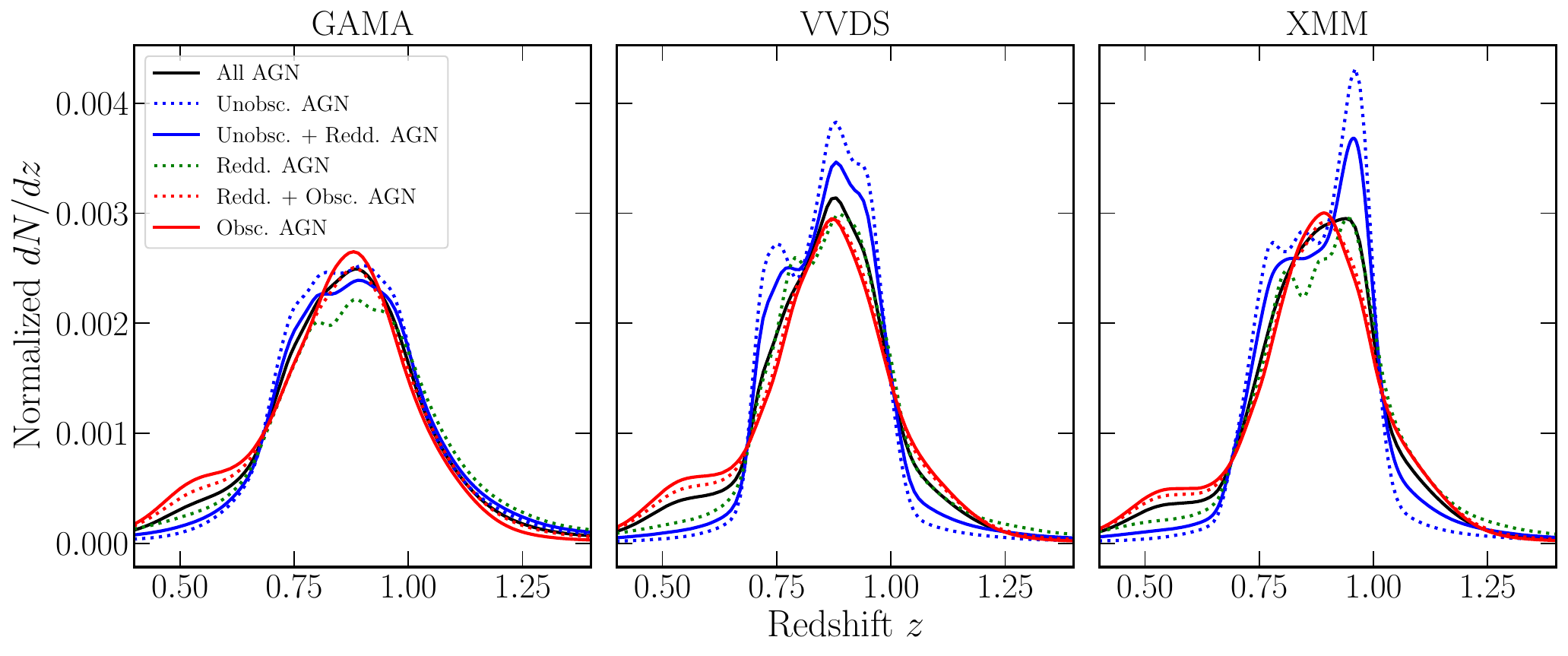}
    \caption{{Normalized $dN/dz$ from all AGN sub-type samples for cross correlation in the wide redshift bin ($z\in 0.7-1.0$) for the $L_{6 \mu m} > 3 \times 10^{44}$ erg s$^{-1}$ threshold, split by HSC field.}}
    \label{fig:all_dNdz}
\end{figure}

{Figure \ref{fig:all_dNdz} highlights the $dN/dz$ for each of the AGN sub-type samples in the principal analysis, for the three HSC fields considered. We find that the AGN $dN/dz$ show multiple peaks and departures from a normal distribution, possibly indicative of fitting degeneracies in the AGN photo-$z$ pipeline. As discussed in \S \ref{sec:model_def}, each of these $dN/dz$ are used to calculate the $\omega(\theta)$ DM model for the particular subset, then used to fit the measured correlation.}

\section{Full Covariance Matrix $\chi^2$ Minimization Fit Results}\label{sec:app_covmat}

\renewcommand{\arraystretch}{1.3}

\begin{table*}
\centering
\caption{ Angular Correlation Function Fit Results (Full Covariance Matrix)} 
\begin{tabular}{cccccccc}
\hline \hline
\multirow{2}{*}{Subset} & \multirow{2}{*}{$N_{obj}$} & \multirow{2}{*}{Weighted $N_{obj}$}  & $\langle L_{6 \mu m} \rangle$ & \multirow{2}{*}{$\langle z \rangle$} & \multirow{1}{*}{$\langle\chi^2\rangle$} & \multirow{2}{*}{$b$} & $\langle M_h \rangle$ \\

 &  &  & [$\log$ erg s$^{-1}$]  & & [10 d.o.f.]   & & [$\log h^{-1} M_\odot$] \\
 \hline \hline
 {$z\in 0.6-0.8 $} & \multicolumn{7}{c}{ } \\
\hline

LRGs & 1,288,589 & 879,258.8 & --   & $0.7\pm0.1$ & 32.8 & $1.83^{+0.04}_{-0.05}$  & -- \\
All AGN & 22,988 & 5,804.3 & $^* 44.7^{+0.4}_{-0.2}$ & $0.8^{+0.1}_{-0.2}$ & 15.8 & $1.7\pm0.1$ & $13.2\pm0.1$ \\
\hline
 {$z\in 0.8-1.0 $} & \multicolumn{7}{c}{ } \\
\hline
LRGs & 851,117 & 440,970.6 & --   & $0.9\pm0.1$ & 28.1 & $2.1^{+0.1}_{-0.2}$ & -- \\
All AGN & 26,264 & 10,381.3  & $^* 44.8^{+0.4}_{-0.2}$ & $0.9^{+0.1}_{-0.2}$ & 29.4 &$1.5\pm0.1$ & $12.8\pm0.1$\\
\hline
 {$z\in 1.0-1.2 $} & \multicolumn{7}{c}{ } \\
\hline
LRGs & 324,790 & 98,498.4 & --   & $1.0\pm0.1$ & 21.2 & $2.43^{+0.01}_{-0.03}$ & -- \\

All AGN & 25,235 & 7,204.7 & $^* 44.9^{+0.4}_{-0.3}$  & $1.1\pm0.2$ & 10.3 & $1.6\pm0.1$&  $ 12.9\pm0.1$\\

\hline
 $z\in 0.7-1.0 $ & \multicolumn{7}{c}{ } \\
\hline

 LRGs & 1,509,905 & 843,166.6 & --  & $0.8\pm0.1$ & 19.2 & $1.9\pm0.1$ & -- \\
All AGN & 28,494 & 13,898.8  & $^* 44.8^{+0.4}_{-0.2}$  & $0.9^{+0.1}_{-0.2}$ & 30.7 & $1.4\pm0.2$ &  $ 13.2^{+0.1}_{-0.2}$ \\

Unobscured AGN & 8,266 & 3,942.0  & $^* 44.9^{+0.4}_{-0.2}$ & $0.9^{+0.1}_{-0.2}$ & 22.3 & $2.3\pm0.2$ & $ 13.4\pm0.1$ \\

Unobscured + Reddened AGN & 15,156 & 6,745.71  & $^* 44.8^{+0.4}_{-0.2}$ & $0.9^{+0.1}_{-0.2}$ & 53.3 & $2.1\pm0.2$ & $ 13.2^{+0.1}_{-0.2}$ \\

Reddened AGN & 6,890 & 2,803.6  & $^* 44.8^{+0.5}_{-0.2}$ & $0.9^{+0.1}_{-0.2}$ & 16.7 & $1.6\pm0.2$ & $ 13.0^{+0.1}_{-0.2}$ \\

Reddened + Obscured AGN & 19,675 & 9,893.4  & $^* 44.8^{+0.4}_{-0.2}$ & $0.9^{+0.1}_{-0.2}$ & 29.2 & $1.4\pm0.2$ & $ 12.7\pm0.2$ \\

Obscured AGN & 12,785 & 7,089.8  & $^* 44.7^{+0.3}_{-0.2}$ & $0.9\pm{0.2}$ & 36.7 & $1.2\pm0.2$ & $ 12.6^{+0.2}_{-0.3}$ \\

\hline
High $L_{6 \mu m}$ AGN & 7,760 & 2,492.2  & $^\dagger 45.1^{+0.4}_{-0.2}$ & $1.0.^{+0.1}_{-0.2}$ & 10.4 & $1.3\pm0.2$ & $ 12.6^{+0.2}_{-0.3}$ \\
Low $L_{6 \mu m}$ AGN & 20,734 & 11,406.5 & $^\ddagger 44.8^{+0.3}_{-0.2}$ & $0.9^{+0.1}_{-0.2}$ & 23.4 & $1.5\pm0.2$ & $ 13.0^{+0.1}_{-0.2}$  \\

\hline \hline
\end{tabular}

\vspace{0.05in}

\raggedright
\footnotesize 
$^*$ Primary luminous AGN selection ($L_{6\mu m} > 3\times 10^{44}$ erg s$^{-1}$)

$^\dagger$ Higher luminous AGN selection ($L_{6\mu m} > 10^{45}$ erg s$^{-1}$)

$^\ddagger$ Lower luminous AGN selection ($3\times 10^{44} < L_{6\mu m} <  10^{45}$ erg s$^{-1}$)

\label{tab:SummMass_BinW_covmat} 
\end{table*}

{Table \ref{tab:SummMass_BinW_covmat} shows the fit results for the complete experiment when using the full covariance matrix, rather than 1-D uncertainties from the square root of the diagonal of the covariance matrix (as shown in Table \ref{tab:SummMass_BinW}). We describe in \S \ref{sec:methods} how this result incorporates the off-diagonal elements showcasing the bin-to-bin correlations, and thus more accurately represents the precision of our measurements. The off-diagonal elements of the cross-correlations' covariance matrices have amplitudes of order $0.2-0.9$ of the value of the diagonal elements. However, for ease of comparison with other analyses of AGN clustering using $\omega(\theta)$ \citep[c.f.][]{koutoulidis_clustering_2013, koutoulidis_dependence_2018, dipompeo_angular_2014, dipompeo_updated_2016, dipompeo_characteristic_2017, petter_host_2023}, we represent the main results using the diagonal elements alongside the effect on the statistical significance when using the full covariance.}

\section{Per-field Correlation Results (1-D Uncertainties)} 

Table \ref{tab:Supplement_tab} shows the per-field measurements of the galaxy and AGN bias {given the 1-D uncertainties}, as well as the inferred average halo mass. We take these measurements and combine them with an inverse variance weighted mean to produce our nominal results, as shown in Table \ref{tab:SummMass_BinW}.

\renewcommand{\arraystretch}{1.3}

\begin{table*}
\centering
\caption{ Angular Correlation Function Fit Results for Each HSC Field} 
\begin{tabular}{cccccccc}
\hline \hline
\multirow{2}{*}{Subset} & \multirow{2}{*}{Field} & \multirow{2}{*}{$N_{obj}$} & \multirow{2}{*}{Weighted $N_{obj}$}  & \multirow{2}{*}{$\langle z \rangle$} & \multirow{1}{*}{$\chi^2$} & \multirow{2}{*}{$b$} & $\langle M_h \rangle$ \\

 & &  &  &  & [10 d.o.f.]   & & [$\log h^{-1} M_\odot$] \\

\hline \hline
 {$z\in 0.6-0.8 $} & \multicolumn{7}{c}{ } \\
\hline

All AGN & GAMA & 15,494 & 3,877.6 & $0.8^{+0.1}_{-0.3}$ & 5.4 & $1.6\pm0.1$ & $12.9\pm0.1$ \\

All AGN & VVDS & 3,815 & 1,050.9 & $0.8^{+0.1}_{-0.3}$ & 10.4 & $1.3\pm0.2$ & $12.5^{+0.3}_{-0.4}$ \\

All AGN & XMM & 3,679 & 8,75.7 & $0.8^{+0.1}_{-0.3}$ & 2.7 & $1.3\pm0.2$ & $12.5^{+0.3}_{-0.4}$ \\

\hline
 {$z\in 0.8-1.0 $} & \multicolumn{7}{c}{ } \\
\hline

All AGN & GAMA & 17,785 & 6,444.19 & $0.9^{+0.1}_{-0.2}$ & 1.0 &$1.5\pm0.1$ & $12.7\pm0.1$\\

All AGN & VVDS & 4,221 & 1,936.76 & $0.9\pm 0.1$ & 7.3 &$0.8\pm0.2$ & $11.2^{+0.5}_{-0.6}$\\

All AGN & XMM & 4,258 & 2,000.31 & $0.9\pm 0.1$ & 5.8 &$1.7\pm0.2$ & $12.9\pm0.2$\\

\hline
 {$z\in 1.0-1.2 $} & \multicolumn{7}{c}{ } \\
\hline

All AGN & GAMA & 17,110 & 4,618.77 & $1.1\pm0.2$ & 4.9 & $1.4\pm0.1$&  $ 12.4\pm0.2$\\

All AGN & VVDS & 3,907 & 1,200.99 & $1.0\pm0.2$ & 1.0 & $1.5\pm0.2$&  $ 12.5^{+0.3}_{-0.4}$\\

All AGN & XMM & 4,218 & 1,384.91 & $1.0\pm0.2$ & 13.2 & $3.1\pm0.3$&  $ 13.5\pm0.3$\\

\hline
 $z\in 0.7-1.0 $ & \multicolumn{7}{c}{ } \\
\hline

All AGN & GAMA & 19,442 & 8,748.86 & $0.9\pm 0.2$ & 7.6 & $1.8\pm0.1$ &  $ 13.0\pm0.1$ \\

All AGN & VVDS & 4,540 & 2,595.26 & $0.9^{+0.1}_{-0.2}$ & 6.7 & $1.1\pm0.2$ &  $ 12.1^{+0.3}_{-0.5}$ \\

All AGN & XMM & 4,512 & 2,554.68 & $0.9^{+0.1}_{-0.2}$ & 5.9 & $1.7\pm0.2$ &  $ 12.9\pm 0.2$ \\

Unobscured AGN & GAMA & 6,510 & 2,800.22 & $0.9^{+0.1}_{-0.2}$ & 8.9 & $2.2\pm0.2$ & $ 13.2\pm0.1$ \\

Unobscured AGN & VVDS & 863 & 590.09 & $0.9\pm0.1$ & 3.0 & $2.0\pm0.3$ & $ 13.1^{+0.2}_{-0.3}$ \\
Unobscured AGN & XMM & 893 & 551.78 & $0.9\pm0.1$ & 9.4 & $2.4\pm0.3$ & $ 13.4^{+0.1}_{-0.2}$ \\

Unobscured + Reddened AGN & GAMA & 11,610 & 4,761.98 & $0.9^{+0.1}_{-0.2}$ & 5.0 & $2.0\pm0.1$ & $ 13.1\pm0.1$ \\
Unobscured + Reddened AGN & VVDS & 1,783 & 1,029.33 & $0.9^{+0.1}_{-0.2}$ & 5.4 & $1.5\pm0.2$ & $ 12.7^{+0.2}_{-0.3}$ \\
Unobscured + Reddened AGN & XMM & 1,763 & 954.39 & $0.9^{+0.1}_{-0.2}$ & 9.9 & $2.1\pm0.2$ & $ 13.2\pm0.2$ \\

Reddened AGN & GAMA & 5,100 & 1,961.77 & $1.0^{+0.1}_{-0.2}$ & 2.7 & $1.6\pm0.2$ & $ 12.9\pm0.2$ \\

Reddened AGN & VVDS & 920 & 439.24 & $0.9\pm 0.1$ & 3.1 & $1.6\pm0.3$ & $ 12.8^{+0.3}_{-0.4}$ \\

Reddened AGN & XMM & 870 & 402.62 & $0.9\pm 0.1$ & 8.2 & $1.7^{+0.3}_{-0.4}$ & $ 12.9^{+0.3}_{-0.5}$ \\

Reddened + Obscured AGN & GAMA & 12,491 & 5,897.68 & $0.9\pm0.2$ & 5.0 & $1.6\pm0.1$ & $ 12.8^{+0.1}_{-0.2}$ \\

Reddened + Obscured AGN & VVDS & 3,617 & 1,998.54 & $0.9^{+0.1}_{-0.2}$ & 8.2 & $1.0\pm0.2$ & $ 11.8^{+0.4}_{-0.6}$ \\

Reddened + Obscured AGN & XMM & 3,567 & 1,997.2 & $0.9^{+0.1}_{-0.2}$ & 4.2 & $1.4\pm0.2$ & $ 12.6^{+0.2}_{-0.3}$ \\

Obscured  AGN & GAMA & 7,391 & 3,935.91  & $0.9\pm{0.2}$ & 6.4 & $1.5\pm0.1$ & $ 12.7\pm0.2$ \\

Obscured  AGN & VVDS & 2,697 & 1,559.3  & $0.9\pm{0.2}$ & 9.8 & $1.0\pm0.2$ & $ 11.8^{+0.4}_{-0.6}$ \\

Obscured  AGN & XMM & 2,697 & 1,594.58  & $0.9^{+0.1}_{-0.2}$ & 6.9 & $1.4\pm0.2$ & $ 12.6^{+0.2}_{-0.3}$ \\

\hline

High $L_{6 \mu m}$ AGN & GAMA & 5,796 & 1,631.12 & $1.0^{+0.2}_{-0.1}$ & 4.5 & $1.3\pm0.2$ & $ 12.4^{+0.3}_{-0.4}$ \\

High $L_{6 \mu m}$ AGN & VVDS & 1,153 & 514.76 & $0.9\pm0.1$ & 3.6 & $1.3^{+0.3}_{-0.4}$ & $ 12.5^{+0.4}_{-0.7}$ \\

High $L_{6 \mu m}$ AGN & XMM & 811 & 346.38 & $1.0\pm0.1$ & 2.9 & $1.7^{+0.3}_{-0.4}$ & $ 12.8^{+0.3}_{-0.5}$ \\

Low $L_{6 \mu m}$ AGN & GAMA & 13,646 & 7,117.74 & $0.9^{+0.1}_{-0.2}$ & 7.4 & $1.8\pm0.1$ & $ 13.0\pm0.1$  \\

Low $L_{6 \mu m}$ AGN & VVDS & 3,387 & 2,080.5 & $0.9^{+0.1}_{-0.2}$ & 5.9 & $1.3\pm0.2$ & $ 12.4^{+0.3}_{-0.4}$  \\

Low $L_{6 \mu m}$ AGN & XMM & 3,701 & 2,208.31 & $0.9^{+0.1}_{-0.2}$ & 6.0 & $1.7\pm0.2$ & $ 12.9\pm0.2$  \\

\hline \hline
\end{tabular}

\label{tab:Supplement_tab} 
\end{table*}

\end{document}